\documentclass[twoside,twocolumn]{iagproc}

\UseRawInputEncoding

\usepackage{booktabs}
\usepackage{multirow}

\usepackage{natbib}
\usepackage{graphicx}
\graphicspath{{figures/}}
\usepackage{eso-pic}
\usepackage{wrapfig}
\usepackage[footnotesize]{caption2}
\usepackage{amsmath}
\usepackage{mathptmx}
\usepackage{indentfirst}
\usepackage{float}
\usepackage{multirow}
\usepackage {url}
\usepackage{blindtext}

\usepackage[
            pdfstartview=FitH,
            CJKbookmarks=true,
            bookmarksnumbered=true,
            bookmarksopen=true,
            colorlinks=true, 
            pdfborder=001,   
            citecolor=blue,            
            linkcolor=blue,          
            ]{hyperref}    

\begin{document}

\pagenumbering{arabic} 

\twocolumn[ \noindent {\footnotesize }\\ %Submitted to JGR, May 2023

\vspace{3mm}

{\Large \noindent {\bf Global geoid model GGM2022}} \\

\noindent{\large WenBin Shen$^{1,2,*}$ 
Youchao Xie$^{1}$, Jiancheng Han$^{3}$, Jiancheng Li $^{1,4}$}

\noindent {\small $^1$ Department of Geophysics, School of Geodesy
and Geomatics, Wuhan University, Wuhan, China \\
$^2$ State Key Laboratory of Information Engineering in Surveying, Mapping and Remote Sensing, Wuhan University, Wuhan 430079, China\\
 $^3$ Institute of Geophysics, China Seismology Bureau, Beijing,China\\
$^4$ Central South University,Changsha, China\\
Corresponding author: W. B. Shen (wbshen@sgg.whu.edu.cn)} \\]

{\small \noindent {\bf{Abstract}}: We provide an updated 5 $^\prime $ $\times $ 5 $^\prime $  global geoid model GGM2022, which is determined based on the shallow layer method (Shen method). First, we choose an inner surface $\Gamma$ below the EGM2008 global geoid by 15 m, and the layer bounded by the inner surface $\Gamma$ and the Earth's geographical surface $S$ is referred to as the shallow layer. The Earth's geographical surface $S$ is determined by the digital topographic model DTM2006.0 combining with the DNSC2008 mean sea surface. Second, we formulate the 3D shallow mass layer model using the refined 5 $^\prime $ $\times $ 5 $^\prime $ crust density model CRUST$_-$re , which { is an improved 5 $^\prime $ $\times $ 5 $^\prime $  density model of the CRUST2.0 or CRUST1.0 with taking into account the corrections of the areas covered by ice sheets and the land-ocean crossing regions. Third, based on the shallow mass layer model and the gravity field EGM2008 that is defined in the region outside the Earth's geographical surface $S$, we determine the gravity field model  EGM2008s that is defined in the whole region outside the inner surface $\Gamma$, where the definition domain of the gravity field is extended from the domain outside $S$ to the domain outside $\Gamma$. Fourth, based on the gravity field EGM2008s and the geodetic equation $W(P)=W_0$ (where $W_0$ is the geopotential constant on the geoid and $P$ is the point on the geoid $G$), we determine a 5 $^\prime $ $\times $ 5 $^\prime $  global geoid, which is referred to as GGM2022. Comparisons show that the GGM2022 fits the globally available GPS/leveling data better than EGM2008 global geoid in the USA, Europe and the western part of China. 

\vspace{2mm}

}\\
\noindent\rule{74mm}{0.6pt}

\vspace{2mm}

\section{Introduction}\label{Introduction}
The determination of the global geoid with centimeter level is a challenge task in physical geodesy in this century. The gravity field model EGM2008 \citep{{Pavlis-etal2008}, {Pavlis-etal2012}} as well as EIGEN-6C4 \citep{Foerste-et-al2014} provides a global $5'\times 5'$ gravity reference frame with the accuracy level of tens of  centimeters. CRUST models\,\citep{{Bassin-etal2000}, Laske2011}\,provides density information in more details than the constant density of 2.67 g/cm$^3$. The digital terrain/elevation model (DTM/DEM, e.g., DTM2006.0, Shuttle Radar Topography Mission)\,\citep{{Pavlis-etal2007}, {Farr-etal2007}} provides the Earth's surface information with ultra-high resolution and relatively high accuracy. All of these datasets (models) provide an opportunity to more precisely determine a global geoid based on a new formulation \citep{Shen2006}.

By general definition, the geoid is the equi-geopotential surface that is nearest to the non-tide mean sea level surface\,\citep{{listing1872},{Heiskanen-and-Moritz1967},{Moritz1980},{Grafarend1994}}. To determine the geoid, a rotational ellipsoid (or reference ellipsoid) is introduced, e.g., the World Geodetic System 1984 (WGS\,84) reference ellipsoid\,\citep{wgs84}. If the geoidal height or geoid undulation $N$ (the distance between the geoid and the ellipsoidal surface along the normal gravity plumb line) is determined, the geoid is determined. The datum of the geoid could be chosen in agreement with international convention \citep{Petit-and-Luzum2010} under the requirement that the determined geoid should be most closed to the non-tide mean sea level. 

Taking the geoid as the boundary, using gravity anomalies on the geoid, based on Stokes approach\,\citep{Stokes1849}, one could determine the disturbing potential field $T(P)$ defined outside the geoid, and by Bruns formula\,\citep{Bruns1878}\,the geoid undulation $N$ could be determined by following expression\,\citep{Heiskanen-and-Moritz1967}
\begin{eqnarray}
\label{Bruns-eq-Stokes} N= \frac{T}{\gamma}
\end{eqnarray}
where $T$ takes the value on the geoid and $\gamma$ is the normal gravity on the surface of the ellipsoid. As a remark, we should keep in mind that equation  (\ref{Bruns-eq-Stokes}) is accurate to first order approximation, derived out based on Taylor expansion \citep{Shen2013a}.

In Section \ref{Serious-considerations} we discuss difficulties existing in Stokes approach and Molodensky approach, some of which are well known, and some of which are seldom known.  In Section \ref{Theoretical-foundation} we provide theoretical foundation for our further formulations in this study. Then, in Section \ref{Shallow-layer-method}, we provide an extended and revised version of \cite{Shen2006} method (might be simply referred to as Shen method), which is difficult to obtain in public community. The Shen method is also referred to as the shallow-layer method \citep{Shen-and-Han2013} for the reasons given therein. Section \ref{Data-and-shallow-layer-geometry}  describes the data sets used, and Section \ref{Shallow-mass-layer-model} describes how to formulate the shallow mass layer model which is significant for determining gravimetric geoid. Section \ref{Determination-of-the-gravitational-potential} describes how to calculate the potential generated by the shallow mass layer, and Section \ref{Global geoid model  2022 (GGM2022)} provides practical computation procedures in details for determining geoid. Section  \ref{Evaluation of the GGM2022}  provides the calculated $5^\prime \times 5^\prime$ global geoid model 2022 (GGM2022) and its evaluation using global available GPS/leveling benchmark data. Finally, conclusions and discussions are given in Section \ref{Conclusions-and-Discussions}.

\section{Considerations about Stokes' and Molodensky approaches}\label{Serious-considerations}

In Stokes approach, to calculate the disturbing potential $T$, it is required the gravity anomaly $\Delta g$ over the whole geoid. Hence, the gravity observations on the Earth's surface or above it should be reduced to the geoid, which requires two conditions: (i) the orthometric height $H$ (the height above the geoid) should be known; (ii) the masses outside the geoid should be moved on or into the inside of the geoid, referred to as to the well-known mass adjustment procedure.  If the external masses (with respect to the geoid) are moved on a surface 21 km below the geoid (in a form of   surface layer), it is referred to first Helmert condensation method\,\citep{{Helmert1884},{Heiskanen-and-Moritz1967}}; and if they are moved just on the geoid (forming a surface layer on the geoid), it is referred to the second Helmert condensation method\,\citep{{Helmert1884},{Lambert1930},{Heiskanen-and-Moritz1967}}. Concerning the second Helmert condensation method, we suggest that the condensed surface layer should locate just below the geoid to avoid theoretical difficulties. For instance, if the condensed layer is just on the geoid, when the field point $P$ goes from geoid to the external space or vice versa, the corresponding gravitation $\nabla V(P)$ (where $V(P)$ is the corresponding gravitational potential) at field point $P$ does not continue\, \citep{Kellogg1929}, giving rise to computation problem in theory, discussion of which in details are beyond this paper.   

Concerning the condition (ii) in Stokes approach, generally, the mass adjustment will change the geoid. Then, one needs further corrections\,\citep{{Heiskanen-and-Moritz1967},{Shen2013a}}. We note that in the case that the Rudzki approach (see e.g., \cite{Heiskanen-and-Moritz1967}) is used, the geoid will not be changed. In Rudzki approach, the masses outside the geoid are moved inside the geoid with complete mirror way, as shown by Figure \ref{Rudzki-mirror}. Though Rudzki approach will change the gravity field outside the Earth, for the purpose of geoid determination, it might be a good method. Unfortunately, Rudzki approach is not considered by present public community. One of the main reasons might be due to the fact that by this method, the geoid should be a priori given; however, it is to be determined. This is of course also the drawback of Stokes' method. Another reason might be due to the fact  that it is complicated in practical calculation. We need iteration procedures, and by every procedure we need new calculations. The third reason could be due to the following fact. Locally examining the Rudzki approach, it will not change the geoid. However, globally examining this approach, it is quite complicated to hold the geoid. Referring to Figure \ref{Rudzki-sphere-mirror}, suppose we have a uniform spherical shell with its inner and outer radii $R_1$ and $R_2$, respectively. Suppose the $R_1$-sphere is a geoid. Now we remove the mass outside $R_1$-sphere into $R_1$-sphere by Rudzki approach. Then, the mirror density distribution $\rho'$ of the mirror shell $R_1 - R_1'$ is not equal to the original density distribution $\rho$ of the shell $R_2$-$R_1$. Obviously, $\rho'>\rho$. 

Figure \ref{Rudzki-global-mirror} shows a real and more complicated case. We cannot simply take the mirror density distribution, which will not hold the geoid invarible. In fact we should carefully adjust the mirror density distribution so that the geoid is not been changed. This might be the key reason why scientists do not use Rudzki approach.  

\begin{figure}[!ht]
  
   \centering
  \includegraphics[scale=0.35]{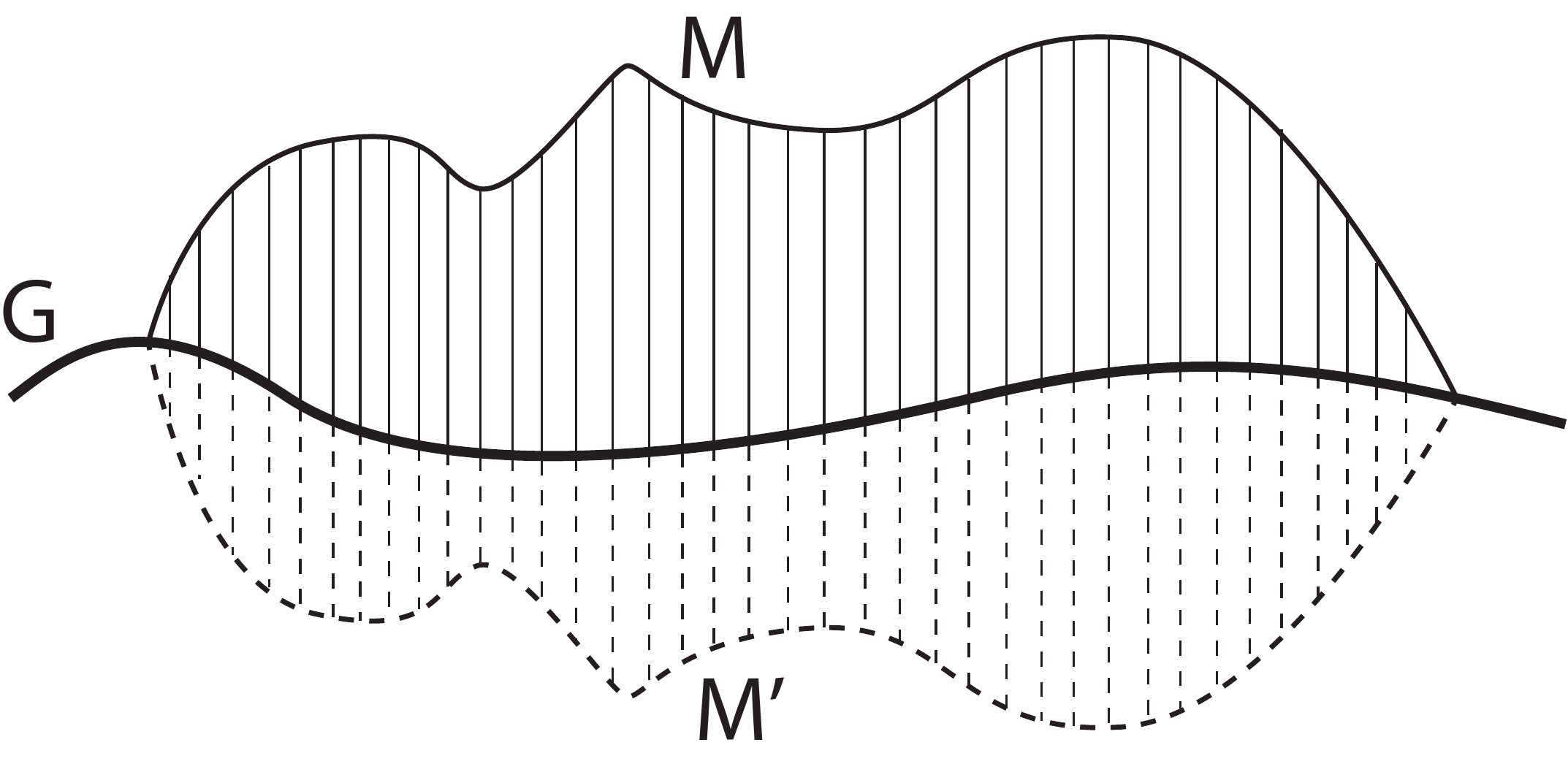}\\
  \caption{{\footnotesize
A scheme of Rudzki approach. The thick solid curve is geoid $G$. The region filled by solid thin segments denotes the mass (mountains) $M$ above the geoid, and the region filled by the dashed segments denotes the mirror mass $M'$ of $M$. By Rudzki approach, $M$ should be removed and put inversely as shown by $M'$.   
 }}\label{Rudzki-mirror}
\end{figure}

\begin{figure}[!ht]
  
   \centering
  \includegraphics[scale=0.3]{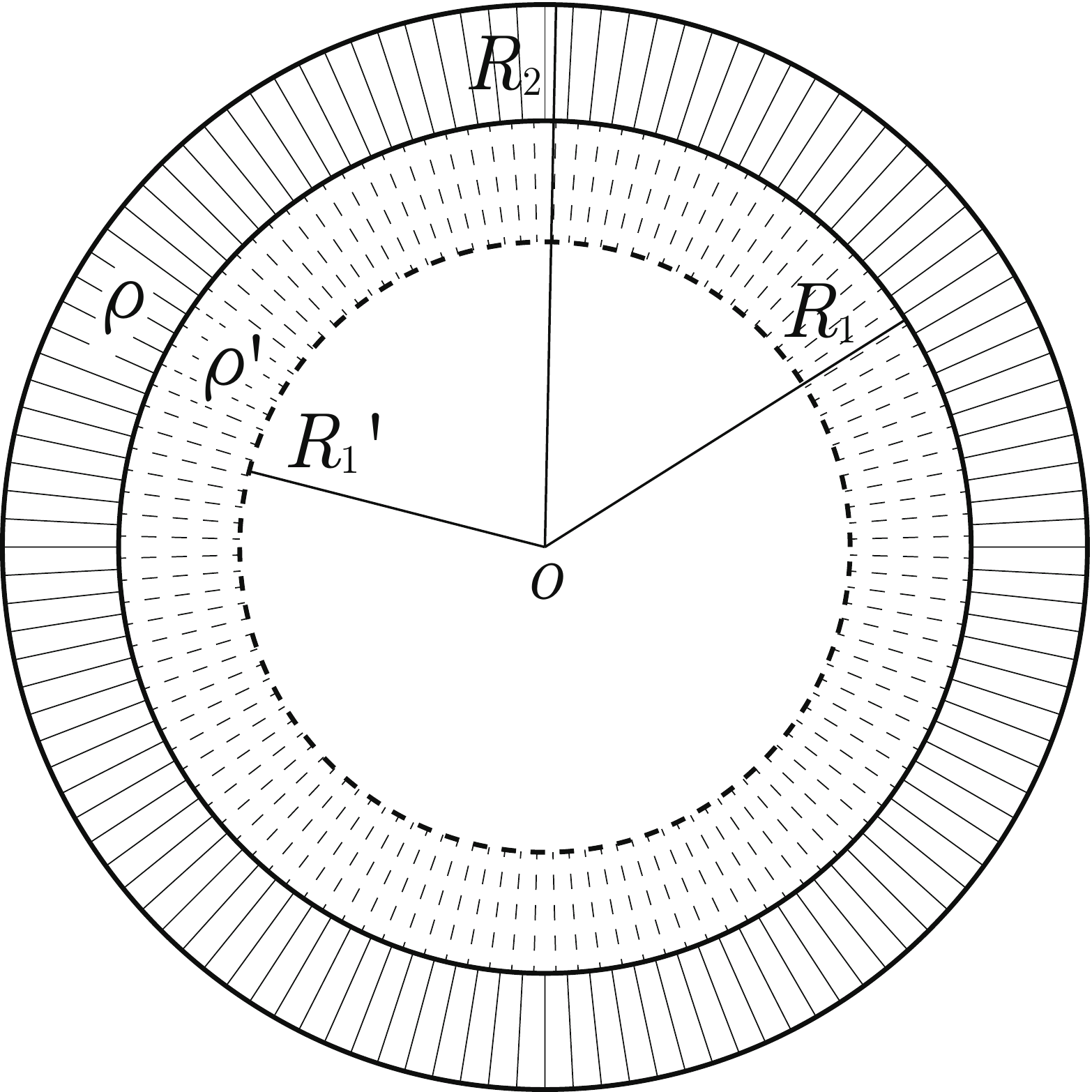}\\
  \caption{{\footnotesize
A scheme of spherical mirror. The thick solid curve is geoid $G$. The region filled by solid thin segments denotes the mass (mountains) $M$ above the geoid, and the region filled by the dashed segments denotes the mirror mass $M'$ of $M$. By Rudzki approach, $M$ should be removed and put inversely as shown by  $M'$.   
 }}\label{Rudzki-sphere-mirror}
\end{figure}

\begin{figure}[!ht]
  
   \centering
  \includegraphics[scale=0.35]{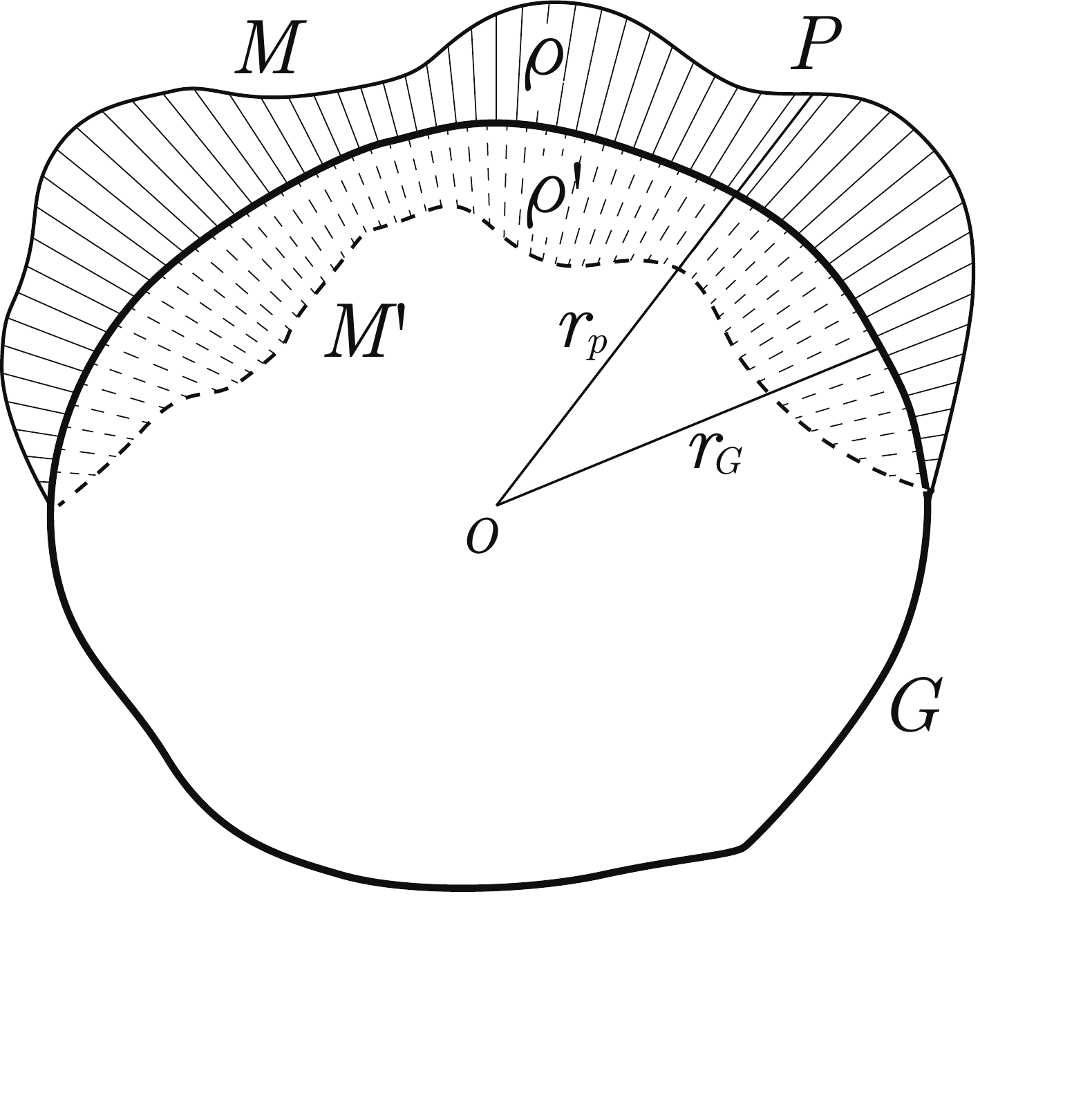}\\
  \caption{{\footnotesize
A scheme of Rudzki mirror with respect to global geoid. The thick solid curve is geoid $G$. The region filled by solid thin segments denotes the mass (mountains) $M$ above the geoid, and the region filled by the dashed segments denotes the mirror mass $M'$ of $M$. By Rudzki approach, $M$ should be removed and put inversely as shown by $M'$.   
 }}\label{Rudzki-global-mirror}
\end{figure}

In another aspect, concerning the condition (i), in the  mountainous areas or at the positions where sequential  leveling transportations for long distances are needed, there exist obvious accumulation errors. Hence, it is quite difficult to guarantee a high accuracy-level (say centimeter-level)  global geoid. The most serious problem lies in that there exists dilemma\, \citep{Shen2013a}: to make mass adjustment, one needs first know the position of the geoid. However, if it were known, one needs not determine the geoid at all. 

Molodensky proposed an approach\, \citep{Molodensky-etal1962} (referred to as Molodensy method or approach) to overcome the drawback of the Stokes method. Based on Molodensky method, taking the Earth's surface as the boundary, using gravity anomaly $\Delta g$ on the Earth's surface, the disturbing potential $T=W(P)-U(P)$ could be determined, and using Bruns formula one could determine the height anomaly of the field point or ground point \, \citep{Moritz1980}
\begin{eqnarray}
\label{Bruns-eq-Mol} \zeta = \frac{T}{\gamma}
\end{eqnarray}
which is the distance between the field point (or ground point) and the telluroid, or the distance between the reference ellipsoidal surface and the quasi-geoid, along the normal gravity plumb line, $\gamma$ is the normal gravity at the corresponding point on the telluroid \citep{{Heiskanen-and-Moritz1967},{Moritz1980}}. Similar to equation (\ref{Bruns-eq-Stokes}), we note that equation (\ref{Bruns-eq-Mol}) is accurate to first order approximation, derived out based on Taylor expansion \citep{Shen2013a}. 

With Molodensky method, the mass adjustment problem existing in Stokes theory is conquered\, \citep{{Molodensky-etal1962}, {Moritz1980}}. However, in Molodenky frame, the determined entity is the quasi-geoid, not the geoid.

As is well-known, the geoid is an equi-geopotential surface and has important applications. The quasi-geoid defined based on the height anomaly $\zeta$ is not an equi-geopotential surface, and consequently it could not be properly and effectively used in practice. To transfer the height anomaly $\zeta $ to geoid undulation $N$, the following equation is used\,\citep{{Heiskanen-and-Moritz1967},{Moritz1980}}
\begin{eqnarray}
\label{N-zeta-eq} N=\zeta + (H^*-H)
\end{eqnarray}
where $H^*$ is the normal height, the height above the quasi-geoid (the distance between the field point or ground point and the quasi-geoid along the normal gravity line), which could be determined by the following formula\,\citep{Heiskanen-and-Moritz1967}:
\begin{eqnarray}
\label{H-star-formula}  H^*= \frac{C}{\gamma _0}[1+(1+f+m-2fsin^2\phi)\frac{C}{a \gamma _0}+(\frac{C}{a \gamma _0})^2]
\end{eqnarray}
where $\gamma _0$ is the normal gravity at the corresponding point $Q_0$ on the ellipsoidal surface, $f$ the geometric flatting of the reference ellipsoid,  $\phi$ the latitude, $a$ the average radius of the Earth, $m$ a parameter defined by the following equation
\begin{eqnarray}
\label{m-eq}  \frac{\Omega^2 a}{\gamma_a}= m+\frac{3}{2}m^2
\end{eqnarray}
where $\Omega$ is the Earth's rotation velocity in terrestial quasi-inertial system, $\gamma_a$ is the normal gravity at equator; $C=W_0-W_P$ is the geopotential number, determined by 
\begin{eqnarray}
\label{C-eq}  C=\int_0 ^P gdh
\end{eqnarray}
where $g$ is the gravity scalar and $dh$ is the leveling increment. 

Hence, to determine $N$ based on equation (\ref{N-zeta-eq}), the orthometric height $H$ is still needed, the errors contained in which are generally quite large in  mountainous areas. 

In another aspect, as \cite{Shen2013a} pointed out, besides the mass adjustment problem, conventional methods (especially Stokes' method and Molodensky method) have additional theoretical difficulties. In the sequel, taking Stokes' method as example, we briefly summary the main points. The details are referred to \cite{Shen2013a}. 

\section{Theoretical foundation}\label{Theoretical-foundation}

In Stokes' theory (or Stokes boundary-value problem), although the Earth's external geopotential $W(P)$ is defined in the region outside the Earth (in fact outside the geoid), the normal gravity potential (or Pizzetti normal gravity potential) $U(P)$ is defined only in the region outside the reference ellipsoid (e.g., WGS84 ellipsoid). $U(P)$ has no definition inside the ellipsoid. Hence, the disturbing potential $T(P)=W(P)-U(P)$ is only defined in the region both outside the geoid and the reference ellipsoid\, \citep{Shen2013a}. That means, in the case that geoid locates inside the ellipsoid (see Figure\, \ref{ellipsoid-geoid}), the disturbing potential cannot be used to determine the geoid, because in this case it has no definition between the geoid and ellipsoid, and the Taylor expansion series cannot be applied in this domain, and consequently the Bruns formula cannot be used\, \citep{Shen2013a}.  However, if instead of the normal reference ellipsoid (e.g. WGS84 ellipsoid) we choose a smaller reference ellipsoid which is completely enclosed by both the geoid and the Earth's surface (see Figure\, \ref{inner-ellipsoid}), the above mentioned difficulties existing in Stokes or Molodensky frame could be solved\, \citep{Shen2013a}. This is also the foundation that we may apply the internationally released gravity field model (e.g., EGM2008)\, \citep{Pavlis-etal2012}, which is defined outside the Earth. The basic idea is stated as follows. 

To solve the difficulties as mentioned above, we choose an inner reference ellipsoid $E_i$, with its center coinciding with the Earth's mass center, its major and minor axes smaller than the corresponding ones of the WGS84 by 150m and 147.9m, respectively \citep{Shen2013a}. The purpose  of choosing such a definite smaller ellipsoid is to overcome the difficulties existing in Stokes frame, because in this case the boundary is the geoid, and the ellipsoid should locate inside the geoid. However, to overcome the difficulties existing in Molodensky frame, we need  to choose a further smaller ellipsoid, so that it completely locates inside the Earth. This is the reason why in the above paragraph we stated that the chosen smaller ellipsoid should be completely enclosed by both the geoid and the Earth's surface. Hence, concerning the Stokes approach, we may choose an inner ellipsoid that is enclosed by the geoid, so that the normal potential field is defined outside the inner ellipsoid and consequently in the whole domain outside the geoid; and concerning the Molodensky approach, we may choose an inner ellipsoid that is enclosed by the Earth's surface, so that the normal potential field is defined outside the inner ellipsoid and consequently in the whole domain outside the Earth. 

For convenience we also use the symbol $E_i$ to denote the surface of the inner ellipsoid $E_i$. The inner reference ellipsoid may generate a normal gravity field $U^*(P)$ defined outside the inner ellipsoid, and in this case the mass contained in $E_i$ is equal to the total mass of the Earth \citep{Shen2013a}, and the normal geopotential on the inner ellipsoidal surface is larger than the geopotential constant $W_0$ on the geoid. The most important thing lies in that in the region outside the ellipsoid $E$ the normal geopotential field  $U^*(P)$ generated by the inner ellipsoid $E_i$ coincides with the normal field $U(P)$ generated by the ellipsoid $E$ with sufficient high precision \citep{Shen2013a}. It should be kept in mind that $U(P)$ has no definition inside the ellipsoid $E$. 

Since in the domain outside $E$, $U(P)$ coincides with $U^*(P)$ (within accuracy requirement),  the normal geopotential field  $U$ generated by the ellipsoid can be considered as the field $U^*(P)$ generated by the inner ellipsoid $E_i$. In fact, the expression formula of $U(P)$ can be naturally downward continued to the surface $E_i$, noted as $U^{**}(P)$ \citep{Shen2013a}.  In this case, the normal gravitational part of the potential field $U^{**}(P)$  is harmonic in the whole region outside the inner ellipsoid $E_i$. That means using the simple continuation of the usual normal potential $U(P)$ could give the same results in terms of the estimate of the disturbing potential $T(P)\equiv W(P)-U(P)$. Study of \cite{Shen2013a} shows that, in the accuracy level of 0.3mm, there is no difference between $U^*(P)$ and $U^{**}(P)$ in the domain outside $E_i$. 

Now, suppose the gravitational potential field EGM2008  \citep{Pavlis-etal2012} or EIGEN-6C4 \citep{Foerste-et-al2014} is given, which is defined in the whole region outside the Earth. The corresponding disturbing potential field $T(P)$ can be also considered as defined in the whole region outside the Earth, in the sense as discussed above; or vice versa. This conclusion should be kept in mind in our new formulation for global geoid determination. Or, for different kinds of purposes, after the definition domain of the gravitational potential is extended (see Section \ref{Shallow-layer-method}), definition domain of the corresponding disturbing potential field $T(P)$ could be also extended accordingly, based on the choice of the inner ellipsoid.  
\begin{figure}[!ht]
   \centering
  \includegraphics[scale=0.6]{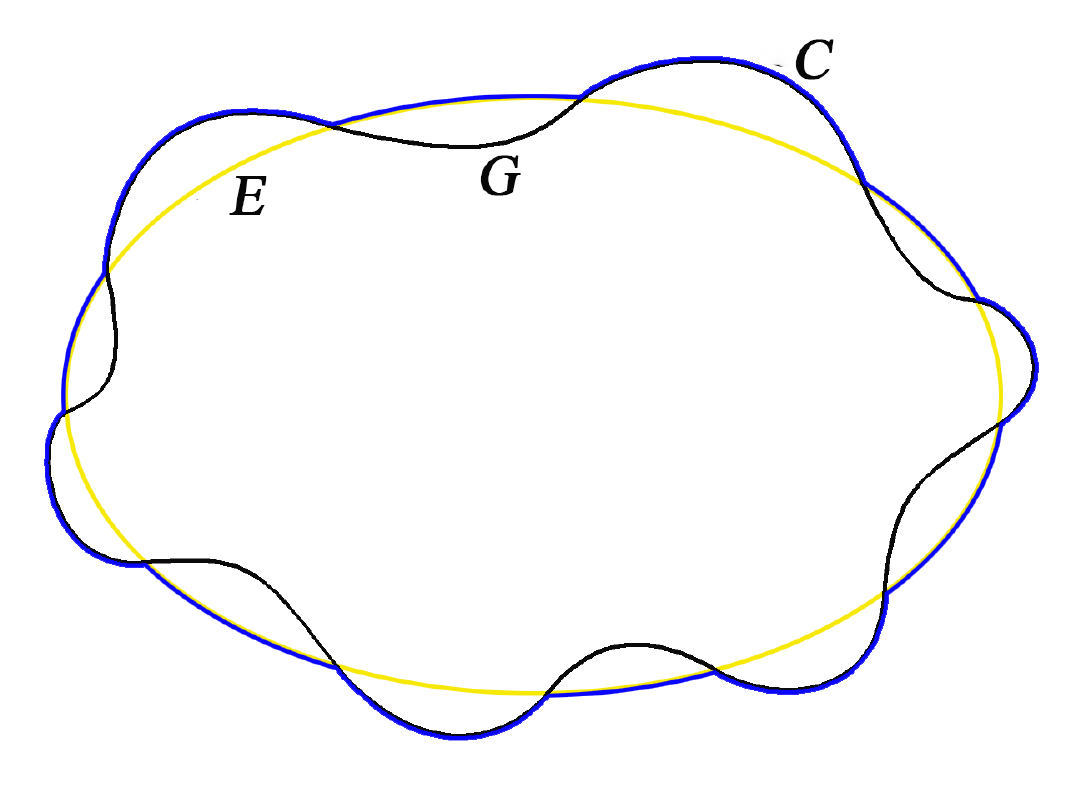}\\
  \caption{{\footnotesize
The blue  line denotes the boundary $C$ that consists of the parts of the geoid that are above the ellipsoidal surface, the parts of the ellipsoidal surface that are above the geoid, and the parts where the geoid coincides with the ellipsoidal surface. The black line denotes the geoid $G$, and the yellow line denotes ellipsoidal surface $E$ {\color{blue} (Modified after \cite{Shen2013a})}. 
 }}\label{ellipsoid-geoid}
\end{figure}

\begin{figure}[!ht]
   \centering
  \includegraphics[scale=0.6]{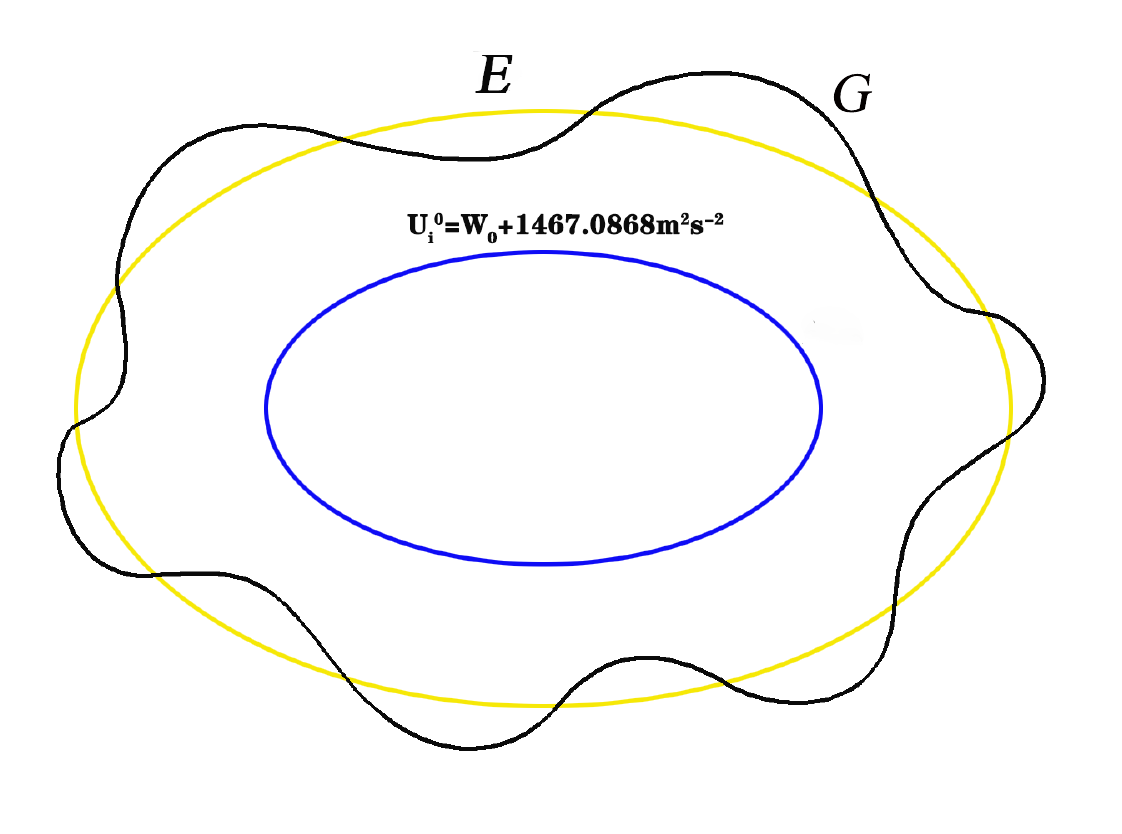}\\
  \caption{{\footnotesize
New formulation of Stokes' approach. The yellow line denotes the surface $E$ of the WGS84 ellipsoid, the black line denotes the geoid $G$, the blue elliptic curve denotes the surface  $E_i$ of the inner ellipsoid. The inner ellipsoid is completely enclosed by the geoid as well as by the Earth's surface {\color{blue} (Modified after \cite{Shen2013a})} 
 }}\label{inner-ellipsoid}
\end{figure}

As a summary, by choosing a smaller ellipsoid, we may overcome the theoretical difficulties existing in Stokes or Molodensky method. In another aspect, it is also a foundation for various approaches for gravity field parameters determinations. Nevertheless, in Stokes method, there exists mass adjustment problem. Though Molodensky method overcomes this problem, it provides a quasi-geoid, the application of which is limited due to the fact that it is not an equi-potential surface. 

Just due to the fact that there exist drawbacks in conventional theories, the original motivation of Shen method \citep{Shen2006} is to solve the following problem: suppose the Earth's surface is provided by high-resolution global digital terrain/elevation model (DTM/DEM, e.g., DTM2006.0, Shuttle Radar Topography Mission) \citep{{Farr-etal2007},{Pavlis-etal2007}}, given a global gravitational potential field model (e.g., EGM2008 \citep{{Pavlis-etal2008},{Pavlis-etal2012}}, or EIGEN-6C4 \citep{Foerste-et-al2014}), and the density distribution model (e.g., CRUST2.0 \citep{Bassin-etal2000}) which provides density distribution information between an inner surface (which lies inside the geoid) and the Earth's surface, the task is to determine a global geoid.

\section{Shallow layer method}\label{Shallow-layer-method}

In this section we briefly outline the shallow layer method (see e.g.,  \cite{Shen-and-Han2013}), or simply Shen method \citep{Shen2006}. Details are referred to relevant publications \cite{{Shen-and-Han2013},{Ashry-etal2021},{Ashry-and-Shen2022},{Xie-etal2021a},{Xie-etal2021b}}. 
%Since the original manuscripts \citep{{Shen2006},{Shen2007}} are difficult to obtain in public community, in subsection \ref{Definition-and-concept-formulation-subsection} we define required entities and concepts, and then in subsection \ref{Formulation-subsection} we provide  the formulation of Shen method for determining a global geoid. 
%In subsection 2.3 we describe the required data sources. In subsection 2.4 we present the way for determining the potential generated by the shallow mass layer, and in subsection 2.5, we discuss relevant computing problems. 

%\subsection{Definition and concept formulation}\label{Definition-and-concept-formulation-subsection}

%To provide the new formulation for determining a global geoid, we first define some required entities used later and describe the data needed in this formulation. 

We constrain our discussions in three dimensional Euclidean space ${\bf E}^3$. Referring to Figure \ref{fig-shallow-layer-def-rev}, the region occupied by the Earth is denoted as  $\Omega_S$, the surface of the Earth is denoted as  $S$, and the region outside the Earth could be expressed as  ${\bf E}^3-\Omega_S \equiv \Omega_S ^c$. For convenience, the geoid is denoted as $G$.  We use $\Omega_G$ to denote the region enclosed by $G$. We choose an inner surface  $\Gamma$, which is completely enclosed by the geoid $G$ (see Figure \ref{fig-shallow-layer-def-rev}),  and we use  $\Omega_\Gamma$ to denote the whole region enclosed by  $\Gamma$, and the region outside the $\Gamma$ could be expressed as  $\Omega_\Gamma ^c$. The layer bounded by the Earth's surface   $S$ and the inner surface  $\Gamma$ is defined as shallow layer. The shallow layer containing the mass is referred to as shallow mass layer. 

For the upper interface, the position of the Earth's surface  $S$ is determined based on the digital elevation model (DEM) on the land. And in the sea, the  $S$ is determined by the DNSC08. For the lower interface, since the variation range of the geoid undulation $N$ is around the magnitude of $\pm$100 meters, generally the depth $D$ of the shallow layer does not exceed 10 km: in ocean areas $D$ is around the magnitude of $\pm$ 2 meters; in land $D$ is generally around $\pm$ 10 to $\pm$ several hundred meters; and in mountain areas $D$ is around several kilometers (e.g., near Everest in Himalaya Mountains $D$ is around 8 km). Geology investigations (including drilling holes) and seismic detection technique can provide more and more precise information about the shallow mass layer. For instance, the preliminary reference Earth model (PREM) \citep{Dziewonski-and-Anderson1981} provides a spherical symmetric density distribution of the Earth, with poor accuracy and without the transverse variation. In the rencent years, there appeared successively more precise density distribution models, CRUST\,5.1 with resolution $5 ^\circ \times 5 ^\circ $ \,\citep{Mooney1998},  CRUST\,2.0 with resolution $2 ^\circ \times 2 ^\circ $ \,\citep{Bassin-etal2000,{Tsoulis2004}} and CRUST\,1.0 \citep{Laske-et-al2013}  with resolution $1 ^\circ \times 1^\circ $ (available at \,{\url{http://igppweb.ucsd.edu/~gabi/crust1.html}}). Hence, the density distribution of the shallow mass layer, denoted by   $\rho_1(P) \,\, (P\in \Omega_S\cap ({\bf E}^3-\Omega_\Gamma)$, namely $P\in \Omega_S\cap \Omega_\Gamma^c)$, is determined based on the crust density model CRUST$_-$re derived from CRUST1.0 or CRUST2.0\citep{Bassin-etal2000,Laske-et-al2013}.

\begin{figure}[!ht]
   \centering
  \includegraphics[scale=0.6]{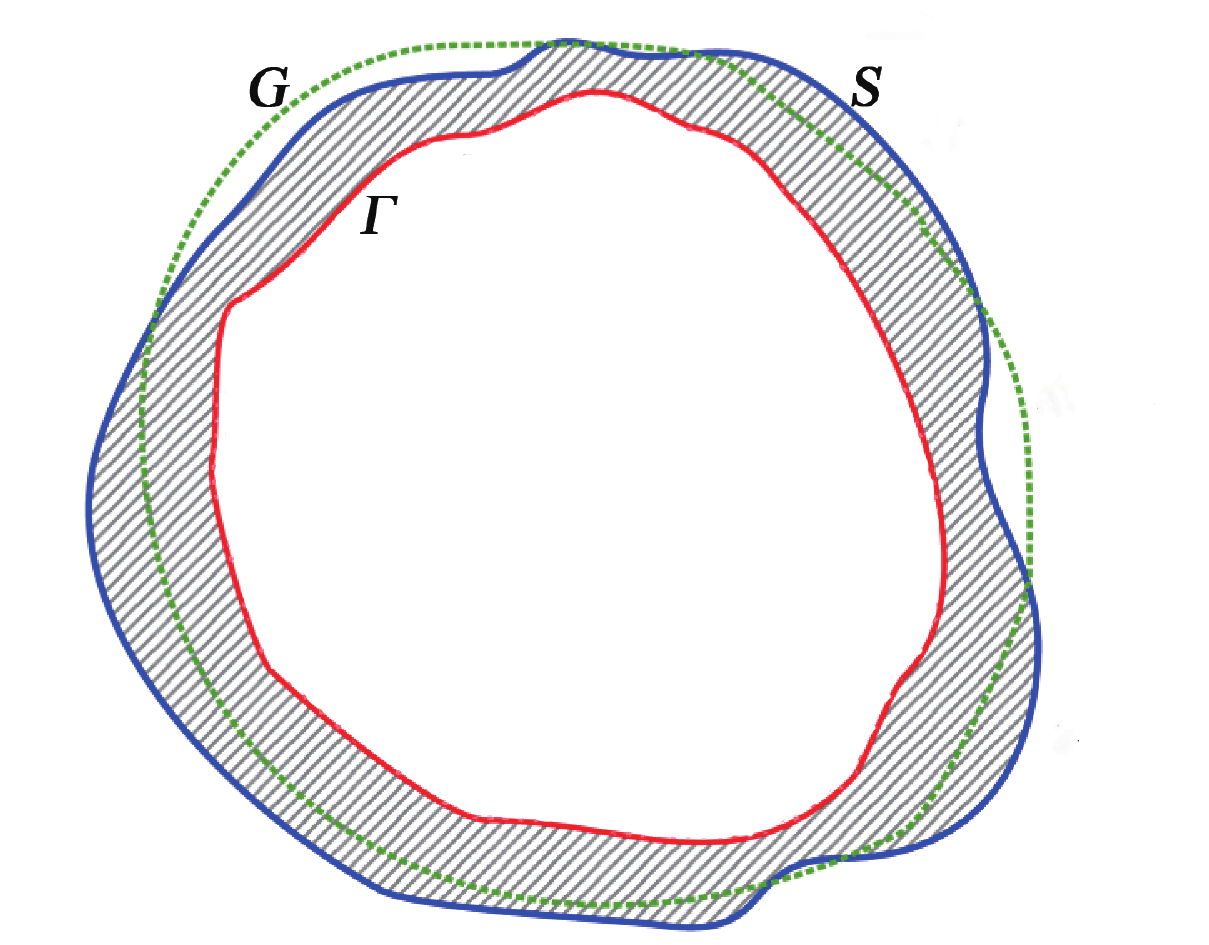}\\
  \caption{{\footnotesize
Definition of the shallow layer, redrawn after \citep{Shen-and-Han2013} . The closed thick solid blue curve denotes the Earth's surface $S$, the closed dotted green curve denotes the geoid $G$, and the thin solid red curve denotes an inner surface $\Gamma$ below the geoid. The masses bounded by $\Gamma$ and $S$ are referred to as the shallow mass layer.  
 }}\label{fig-shallow-layer-def-rev}
\end{figure}

\vspace{3mm}

%\subsection{Formulation}\label{Formulation-subsection}

Due to the construction of the shallow layer, the gravitational potential $V(P)$  of the whole Earth can be divided into two parts. One part is the potential field $V_0(P)$  generated by the mass enclosed by $\Gamma$ (referred to as the inner mass), and another part is the potential field$V_1(P)$  generated by the shallow layer mass. We can obtain the gravitational potential of the two parts by Newtonian integral and natural continuation in the domain outside $\Gamma$. 

The gravitational potential generated by the shallow mass layer is
denoted as $V_1(P)$, which could be determined based on the Newtonian
potential formula
\begin{eqnarray}
\label{shallow-layer-integral-eq}
V_1(P)=G_{grav}\int_{\Omega_\Gamma^c\cap \Omega_S}\frac{\rho_1}{l}d\tau, \,\,
P\in {\bf E}^3
\end{eqnarray}
where $\rho_1$ is the density distribution of the shallow mass layer (which occupies the domain $\Omega_S\cap \Omega_\Gamma^c$), $G_{grav}$ is the gravitational constant;  $l$ is the distance between the field point $P$ and the volume integration element
$d\tau$. It is noted that Eq. (\ref{shallow-layer-integral-eq}) is defined in the whole space ${\bf E}^3$. Once $\rho_1$ is given, $V_1(P)$ can be determined.

Therefore, the gravitational potential of the whole Earth can be expressed as:
\begin{eqnarray}
\label{superposition-principle-V-eq}
V(P)=V_1(P)+ V_0(P), \,\, P\in \Omega_S ^c
\end{eqnarray} 
and the geopotential is expressed as:
\begin{eqnarray} \label{initial-geoid-eq}
W(P)=V(P)+Q(P), \,\, P\in \Omega_S ^c
\end{eqnarray}
where $Q(P)$ is the centrifugal force potential, $V(P)$ is the gravitational potential, $P$ is the field point. Since $Q(P)$ is known, given the gravitational potential field $V(P)$, the  geopotential field $W(P)$ is determined, and vice versa.Therefore, we can determine the geopotential at an arbitrary point in the domain outside $\Gamma$. Finally, determining the position of the geoid precisely is a problem of solving the following equation:
\begin{eqnarray}
\label{geoid-eq} 
V(P)+Q(P)=W_0
\end{eqnarray}
where $W_0$ is the geopotential constant on the geoid, which might be chosen as $W_0=62636851.7146$ m$^2$s$^{-2}$. In practical applications, $W_0$ should be chosen in such a way that the determined geoid is nearest to the (non-tide) mean sea level. In determination of the geoid based on equation (\ref{geoid-eq}), ``iterative method''  could be used. Exactly saying, the following procedures could be applied.

Procedure 1. It is referred to Fig.\ref{omega-geoid-gamma-ray2-rev1}. Introducing a ray $l$ along an
arbitrary direction $(\theta, \lambda)$ from the coordinate origin
$o$ which coincides with the mass center of the Earth, it intersects
a point $P_0$ on a known geoid model, e.g., EGM2008 geoid. Starting from $P_0$,
along the ray $l$ moving slowly toward infinity, there must exists one and only one point $P_G$ so that equation (\ref{geoid-eq}) holds,
because different equi-geopotential surfaces never intersect with
each other. In fact, defining
\begin{eqnarray} \label{f-eq-V-and-Q}
f(P)=V(P)+Q(P)
\end{eqnarray}
and fixing $P$ on the ray $l$, since the direction $(\theta,
\lambda)$ is fixed, $f(P)$ is a monotonous descending function of
$P$ (the discussions are constrained in the near-Earth space).

Procedure 2. First, it might be chosen $P_0$ and let it enter equation
(\ref{geoid-eq}). If $P_0$ is just located on the geoid (i.e.,
$P_0=P_G$), equation (\ref{geoid-eq}) holds; otherwise, it must hold
\begin{eqnarray}
\label{f-lth-Wzero} 
f(P)> W_0
\end{eqnarray}
That means the point $P_G$ should be searched at a further distance from the origin. In fact, one can use Bruns formula to speed up the procedure. Defining the step length
\begin{eqnarray} \label{step-length-def}
l_i=\frac{f(P_{i-1})-W_0}{\gamma},\,\,i=1,2,\cdots,
\end{eqnarray}
then, along the ray $l$, starting from point $P_{i-1}$ a further point $P_i$ could be tested according to the step length $l_i$.

Procedure 3. As the $N$-th step is tested, if
\begin{eqnarray}
\label{limitation-delta} 
|\frac{f(P_N)-W_0}{\gamma} |<\delta
\end{eqnarray}
holds, then stop; otherwise, continue the above procedures. In equation (\ref{limitation-delta}), $\delta$ is a critical value for certain accuracy requirement. For instance, if we need to determine one-centimeter level geoid, then we should set  $\delta < 0.001$ m. In practice, we may set it smaller. 
\begin{figure}[!ht]
   \centering
  \includegraphics[scale=0.6]{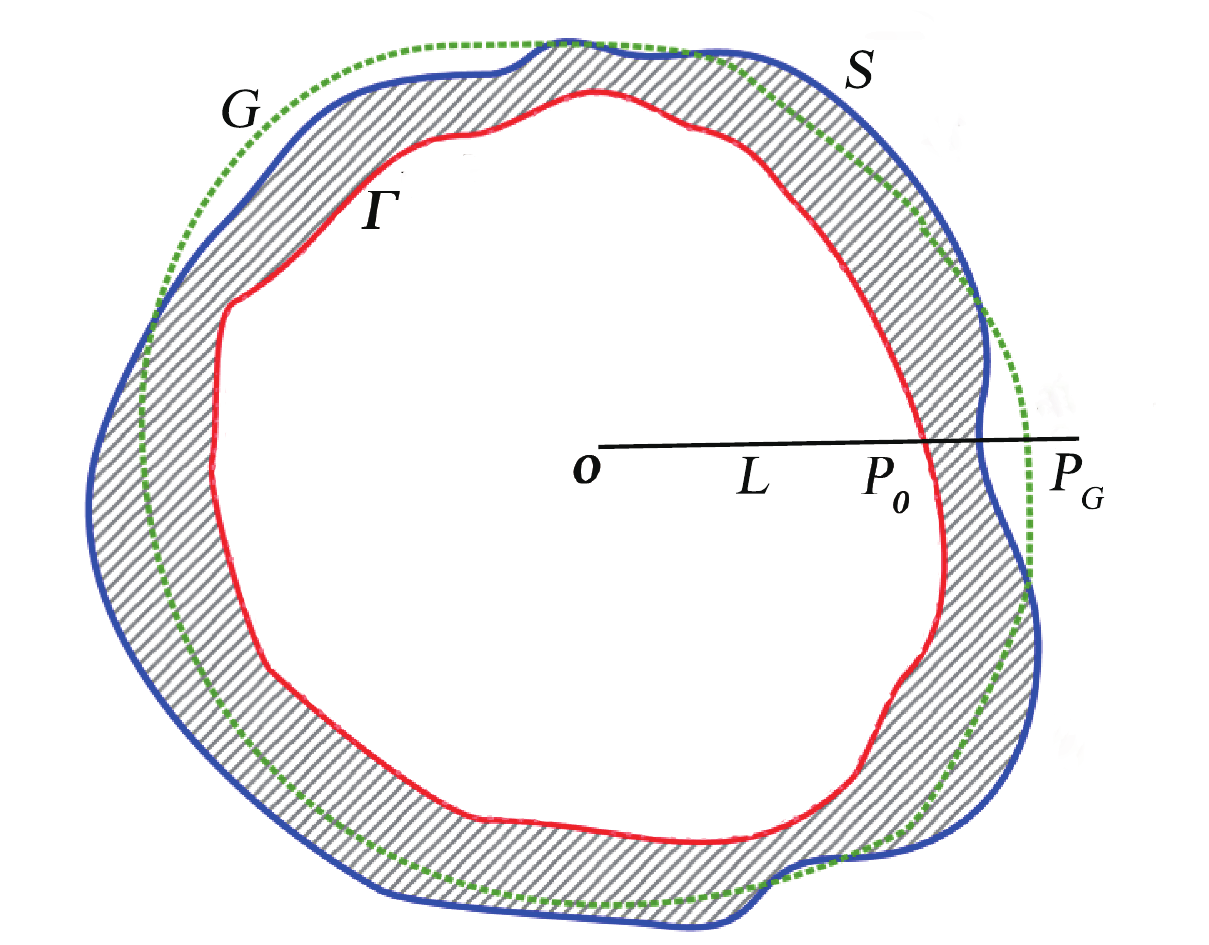}\\
  \caption{{\footnotesize
 The closed thick solid blue curve denotes the Earth's surface $S$, the closed dotted green curve denotes the geoid $G$, and the thin solid red curve denotes an inner surface $\Gamma$ below the geoid. The masses bounded by $\Gamma$ and $S$ are referred to as the shallow mass layer.  The ray line $L$ intersects at points $P_0$ and $P_G$ on the surfaces $\Gamma$ and $S$, respectively (Redrawn after \cite{Han2012}).
 }}\label{omega-geoid-gamma-ray2-rev1}
\end{figure}

\vspace{3mm}

As a relevant remark, concerning the conventional approaches, once the disturbing potential $T(P)$ is given, in plain areas, one can determine the geoid based on the Bruns formula (first-order approximation):
\begin{eqnarray}
\label{Bruns-formula-re} 
N = \frac{T(P)|_{P\in E}}{g(P)|_{P\in
E}}
\end{eqnarray}
where both $T(P)|_{P\in G}$ and $g(P)|_{P\in 
E}$ are known, $E$ is the surface of the reference
ellipsoid, $T(P)=W(P)-U(P)$ is the disturbing potential field,
$U(P)$ is the normal potential field generated by the reference
ellipsoid, or rigousely the reformulated normal potential field $U^*(P)$ generated by an inner reference
ellipsoid \citep{Shen2013a}. Eq. (\ref{Bruns-formula-re}) could be directly derived out based on Taylor expansion, noting that $U(P)|_{P\in E}\equiv W_0$ \citep{{Heiskanen-and-Moritz1967}, {Shen2013a}}. We note that, in mountainous areas, Eq. (\ref{Bruns-formula-re}) should be modified.

\section{Data and shallow layer geometry}\label{Data-and-shallow-layer-geometry}

The $5^\prime \times 5^\prime$ resolution ($\sim 10$ km) geopotential model, EGM2008 \citep{{Pavlis-etal2008},{Pavlis-etal2012}}, is at present the most precise ultrahigh-degree global geopotential model of the Earth's external gravity field (available at website:  {\url{http://earth-info.nga.mil/GandG/wgs84/gravitymod/egm2008/egm08_wgs84.html}}). It is complete to spherical harmonic degree and order 2159, and contains additional spherical harmonic coefficients extending to degree 2190 and order 2159. EGM2008 has been developed by combining the spaceborne GRACE satellite data, terrain and altimetry data, and the surface gravity data \citep{Kenyon-etal2007}. An equivalent external gravitational potential field, namely EIGEN-6C4 \citep{Foerste-et-al2014} (available at website:\, {\url{http://icgem.gfz-potsdam.de/ICGEM/}}), could be also applied if one wishes. Based on Shuttle Radar Topography Mission (SRTM) \citep{Farr-etal2007} data and other altimetry data sets, the high-resolution global digital topographic model DTM2006.0 complete to degree/order 2160 \citep{Pavlis-etal2007} was also publicly available. 

   In order to calculate the gravitational potential of the shallow mass layer, according to Newtonian integral, one has to know {\color{blue} (i) } the geometry of the entire shallow layer, and {\color{blue} (ii) } its interior structure, especially its density distribution. The density distribution is usually provided by geological investigations (rock samples, deep drilling projects, etc.) and active seismic methods. \cite{Dziewonski-and-Anderson1981} presented the preliminary reference Earth model (PREM) with a spherical symmetric density distribution of the Earth. From then on, many other models have been presented with various levels of details. In this study, the available global crustal model, CRUST2.0/CRUST1.0\citep{Bassin-etal2000,Laske-et-al2013} (available at {\url{http://igppweb.ucsd.edu/~gabi/crust1.html}}), are adopted. Evaluations and  analyses of crustal models are given, for instance, by  \cite{Bassin-etal2000} and\, \cite{Laske-et-al2013}. CRUST2.0 offers a detailed density distribution and structure of the crust on a $2^\circ \times 2^\circ$ grid, and defines 360 crustal types, each grid cell consisting of 7 layers: (1) ice, (2) water, (3) soft sediments, (4) hard sediments, (5) upper crust, (6) middle crust, and (7) lower crust. Each $2^\circ \times 2^\circ$ cell is assigned to one kind of crustal type where the compressional and shear wave velocity (V$_P$, V$_S$), density and the upper and lower boundaries are given explicitly for each individual layer. CRUST1.0 defines 35 crustal types and compared with CRUST2.0, in each 1-degree cell of CRUST1.0, boundary depth, compressional and shear velocity as well as density is given for 8 layers: water, ice, 3 sediment layers and upper, middle and lower crystalline crust.
   
The determination of the geometry of the shallow layer is discussed as follows. First, we focus on the upper surface of the shallow layer, namely, the topographic surface. A digital terrain/elevation model (DTM/DEM) with a specific grid resolution can be used to represent the topographic surface. This representation depends on a discretization due to the fact that DTM/DEM is usually given at scattered locations or on geographical grids. For the numerical evaluation in this study, the global digital topographic model DTM2006.0 mentioned earlier is used: this is a model created to supplement EGM2008. DTM2006.0 provides elevation on land areas and bathymetry on ocean areas for an arbitrary point. However, this is not sufficient for our present study. What we need is the topographic surface on both continents and ocean surface. Fortunately, in ocean areas we may use the Danish National Space Center data set DNSC08 mean sea surface (MSS), established from an integration of satellite altimetry data with a time span from 1993 to 2004\, \citep{{Andersen-and-Knudsen2009},{Andersen-etal2010}}. Hence, the upper surface $S$ of the shallow layer is established by combining DTM2006.0 on solid land areas and DNSC08 MSS on ocean surfaces. DTM2006.0 model is available at website ({\url{http://earth-info.nga.mil/GandG/wgs84/gravitymod/egm2008/egm08_wgs84.html}}), see Figure\,\ref{fig-upper-surface-land}; and DNSC08 MSS is available at DTU website ({\url{http://www.space.dtu.dk/English/Research/Scientific_data_and_models/Global_Mean_sea_surface.aspx}}), see Figure\,\ref{fig-upper-surface-ocean}. The whole upper surface consists of the upper surfaces both in solid land and ocean parts.

Second, we have to choose the lower surface of the shallow layer, namely, the inner surface $\Gamma$ (Cf. Figure \ref{fig-shallow-layer-def-rev}). Theoretically, $\Gamma$ can be an arbitrary closed smooth surface that lies inside the geoid $G$ \citep{Shen2006}. Since the geoid undulations vary globally within $\pm$110 m, it is easy to determine an approximate position of the surface $\Gamma$. In order to simplify the description and calculations, generally we may choose the geoid of EGM2008 as a reference surface. Then, a new surface that extends from the reference surface downward to a sufficient depth (e.g., 150 meters or 15 meters) is constructed, which is referred to as the inner surface $\Gamma$ (Cf. Figure \ref{fig-shallow-layer-def-rev}). Here "a sufficient depth" below reference surface means that it should be guaranteed that the inner surface $\Gamma$ should be completely enclosed by the geoid. {\color{blue} In this study, we choose the inner surface $\Gamma$ smaller than (below) the EGM2008 geoid by 15 m}. Now both the upper and lower surfaces of the shallow layer have been determined. Hence, the lower surface $\Gamma$ is constructed by extending from the EGM2008 geoid $G_{EGM08}$ downward to a depth of 15 m. Exactly saying, taking $G_{EGM08}$ as a reference surface, then we choose an inner surface $\Gamma$ that is below the $G_{EGM08}$ by 15 m, see Figure \ref{fig-shallow-layer-def-rev} as reference. Here we note that, by choosing the inner surface $\Gamma$ in this way, it is guaranteed that $\Gamma$ is completely enclosed by the geoid $G$ because it is certain that the largest deviation between $G_{EGM08}$ and the real geoid $G$ {\color{blue} does not exceed} 10 m.    
   
Third, we establish shallow mass layer model, in which the density distribution is assigned to the shallow layer geometric body, see Section \ref{Shallow-mass-layer-model} in details. 

Finally, we use the combination modeling method (CMM) as described in Section \ref{Determination-of-the-gravitational-potential} to calculate the gravitational potential generated by the shallow mass layer, see Section \ref{Global geoid model  2022 (GGM2022)}.

\section{Shallow mass layer model}\label{Shallow-mass-layer-model}

To determine the geoid based on Shen method, we need 3D shallow mass layer model (SMLM). Section \ref{Data-and-shallow-layer-geometry} shows how to construct a geometric shallow layer. In the sequel we show how to construct the SMLM. The key problem lies in that how the density distribution is assigned to the shallow layer body.

\begin{figure}[!ht]
   \centering
  \includegraphics[scale=0.55]{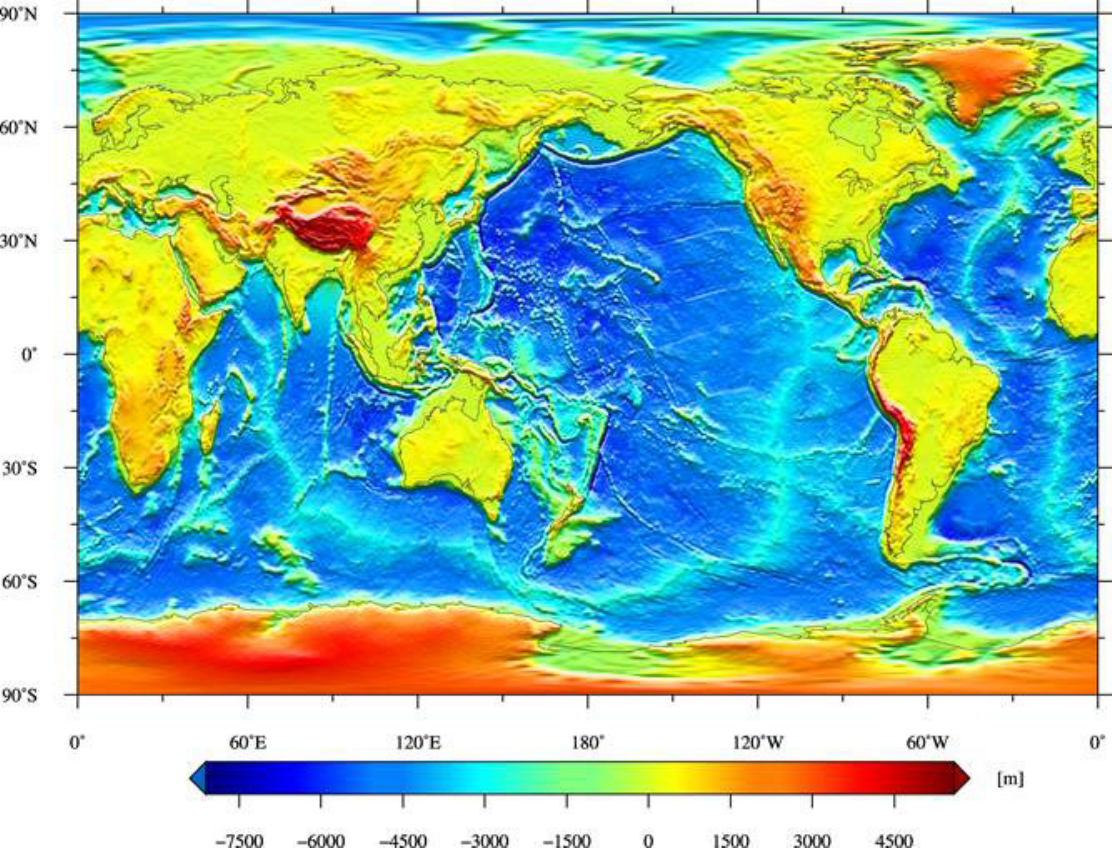}\\
  \caption{{\footnotesize
Upper surface  in land, provided by DTM2006.0 ({\url{http://earth-info.nga.mil/GandG/wgs84/gravitymod/egm2008/egm08_wgs84.html}})
 }}\label{fig-upper-surface-land}
\end{figure}

\vspace{3mm}

\begin{figure}[!ht]
   \centering
  \includegraphics[scale=0.55]{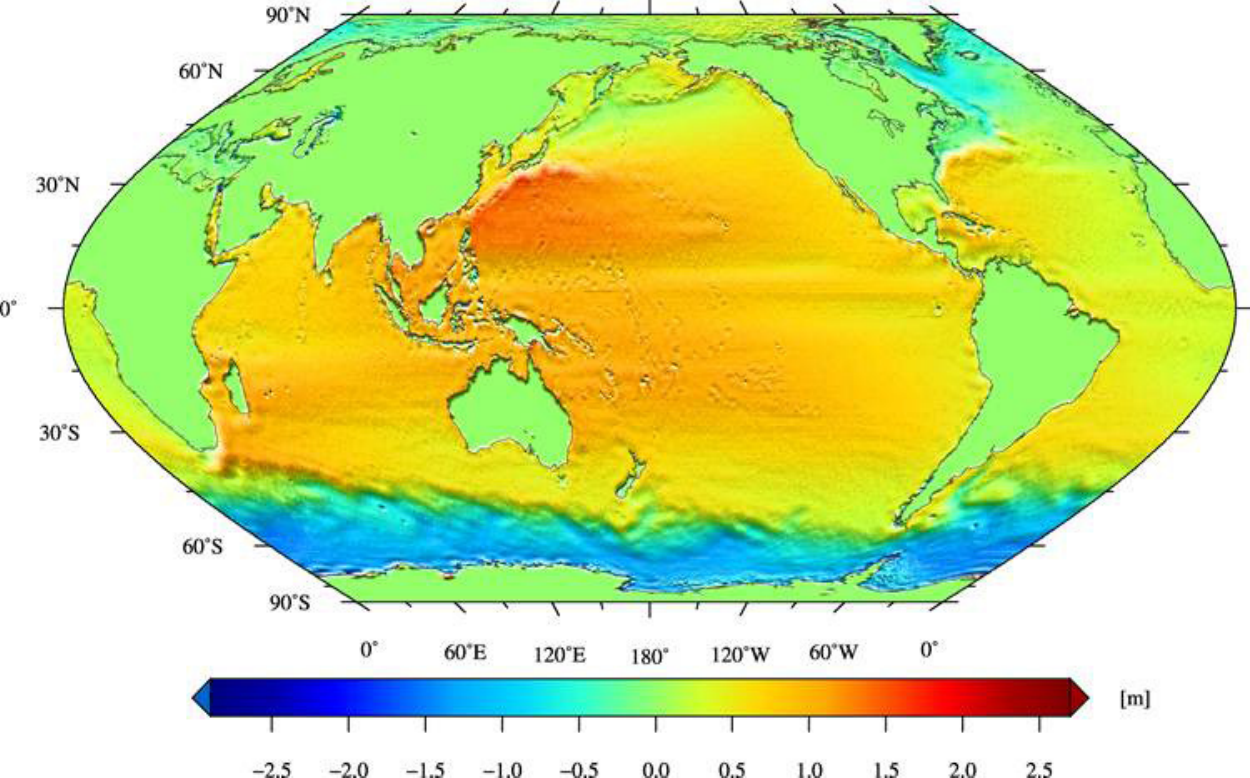}\\
  \caption{{\footnotesize
Upper surface in ocean, provided by DNSC08 ({\url{http://www.space.dtu.dk/English/Research/Scientific_data_and_models/Global_Mean_sea_surface.aspx}})
 }}\label{fig-upper-surface-ocean}
\end{figure}

\vspace{3mm}

The body density distribution inside the shallow layer body is assigned by density information of CRUST$_-$re, which is a refined model based on CRUST2.0/ CRUST1.0  {(as shown by Figure \ref{fig-crust-model-thick}, and available at {\url{http://igppweb.ucsd.edu/~gabi/crust2.html}}). }

\begin{figure}[!ht]
   \centering
  \includegraphics[scale=0.35]{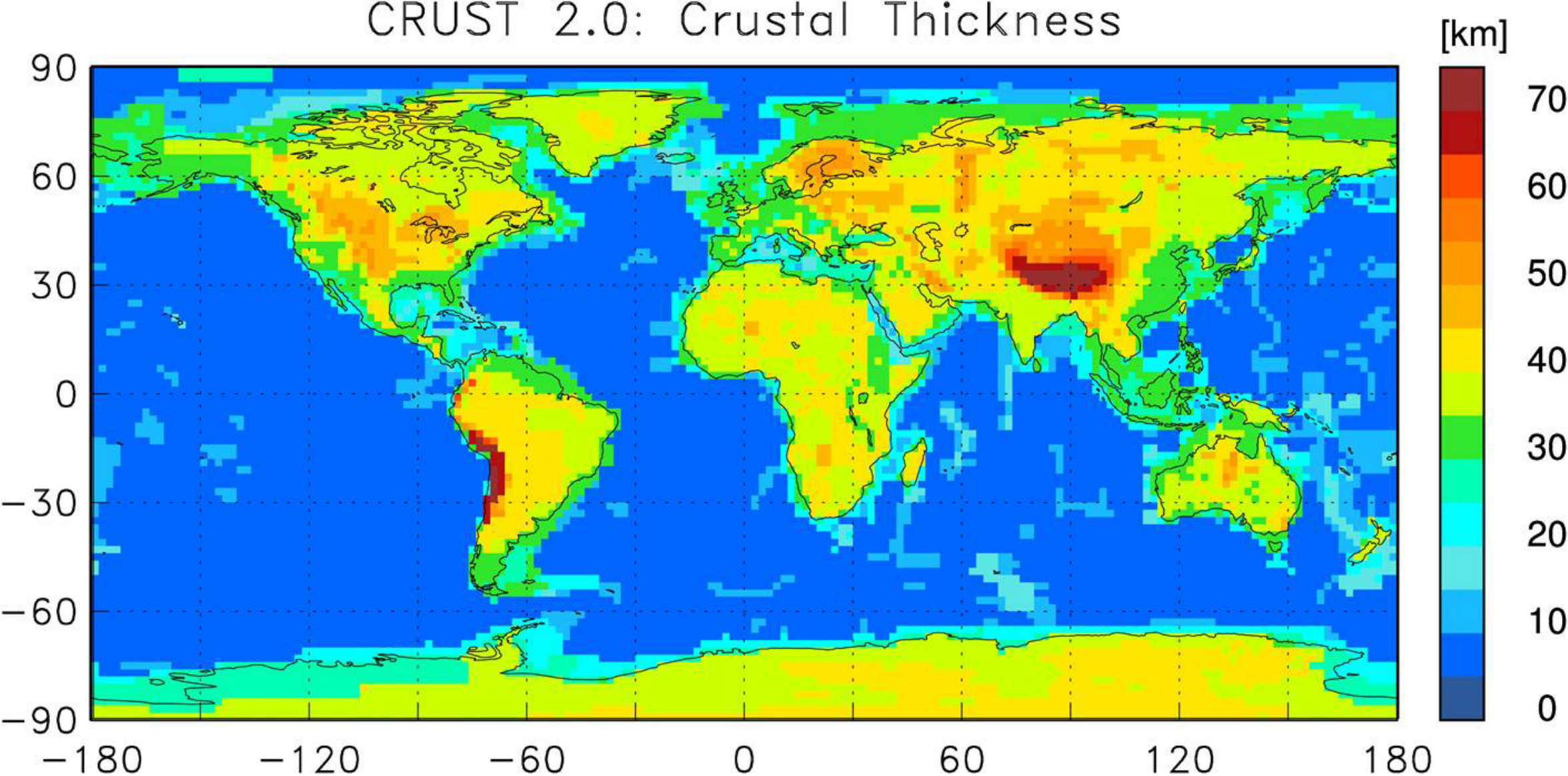}\\
  \caption{{\footnotesize
Thickness information of CRUST2.0 ({\url{http://igppweb.ucsd.edu/~gabi/crust2.html}})
 }}\label{fig-crust-model-thick}
\end{figure}

\vspace{3mm}

CRUST$_-$re is a refined $5^\prime \times 5^\prime$ crust density model based on CRUST2.0/CRUST1.0. In this study, CRUST$_-$re was generated by the following procedures. 

(i) The crust surface is divided into $5^\prime \times 5^\prime$ grid cells.

(ii) Based on ICE-5G model \citep{Peltier2004} we make the ice coverage and thickness corrections. The ice coverage information is shown by Figure\,\ref{fig-ice-coverage-thickness-corr}. We replace the ice coverage and thickness information given by CRUST2.0/CRUST1.0 with those given by ICE-5G model, which is available at website ({\url{http://www.atmosp.physics.utoronto.ca/~peltier/data.php}}).

(iii) On the boundaries between land and ocean, we make corrections, due to the fact that CRUST2.0/CRUST1.0 provides density information only in cell resolution of  $2^\circ \times 2^\circ$ or $1^\circ \times 1^\circ$. Figure\,\ref{fig-ocean-land-boundary-inf} shows the land-ocean boundary information, based on which we correct the density distribution information provided by CRUST models. Figure \ref{fig-ocean-land-boundary-TW} provides an example, showing what happens when a grid includes both ocean area and land area. A $1^\circ \times 1^\circ$ grid (see region $A$ of Figure \ref{fig-ocean-land-boundary-TW}) given by CRUST1.0 contains both ocean and land, but it provides only a unique  density value (e.g., 1.02 $g/cm^3$). However, in fact it contains two parts: one part is water (e.g. with density 1.02 $g/cm^3$), but other part is land (e.g. with density 2.67 $g/cm^3$). Hence, we need to refine the density information.  

For each $5^\prime \times 5^\prime $ grid ($i,j$), we consider the following real situation, see Figure\,\ref{fig-grid-corr}. If a grid given by CRUST1.0 is land, but in real it is ocean, then we change the original given density into $1.02 g/cm^3$. If a grid given by CRUST1.0 is ocean, but in real it is land, then we change the original given density into $2.67 g/cm^3$ (or more practical density value such as $2.10 g/cm^3$, depending on the practical situation).

 In practice, suppose the $ij$-th grid locates partly in land and partly in ocean, we first find the surrounding 8 grids, take the average of the thicknesses of the 8 surrounding grids as the thickness of the $ij$-th grid, and take the average density of the 8 grids as the density of the $ij$-th grid.   In the case that there is lack of thickness information around the $ij$-th grid (It occurs for islands and very seldom), it means it quite small (otherwise it could be informed by DEM), and we may consider it as ocean. 

Generally, in plain land area or ocean area, shallow mass layer contains one or two layers provided by CRUST$_-$re. In mountainous areas (e.g., Tibetan area), shallow mass layer may contain three or more layers. Figure \ref {fig-diff-layers} shows an example in Tibetan area \citep{Han2012}, where the shallow mass layer contains four layers provided by CRUST$_-$re : (1) ice, (3) soft sediments, (4) hard sediments, and (5) upper crust.
\begin{figure}[!ht]
   \centering
  \includegraphics[scale=0.42]{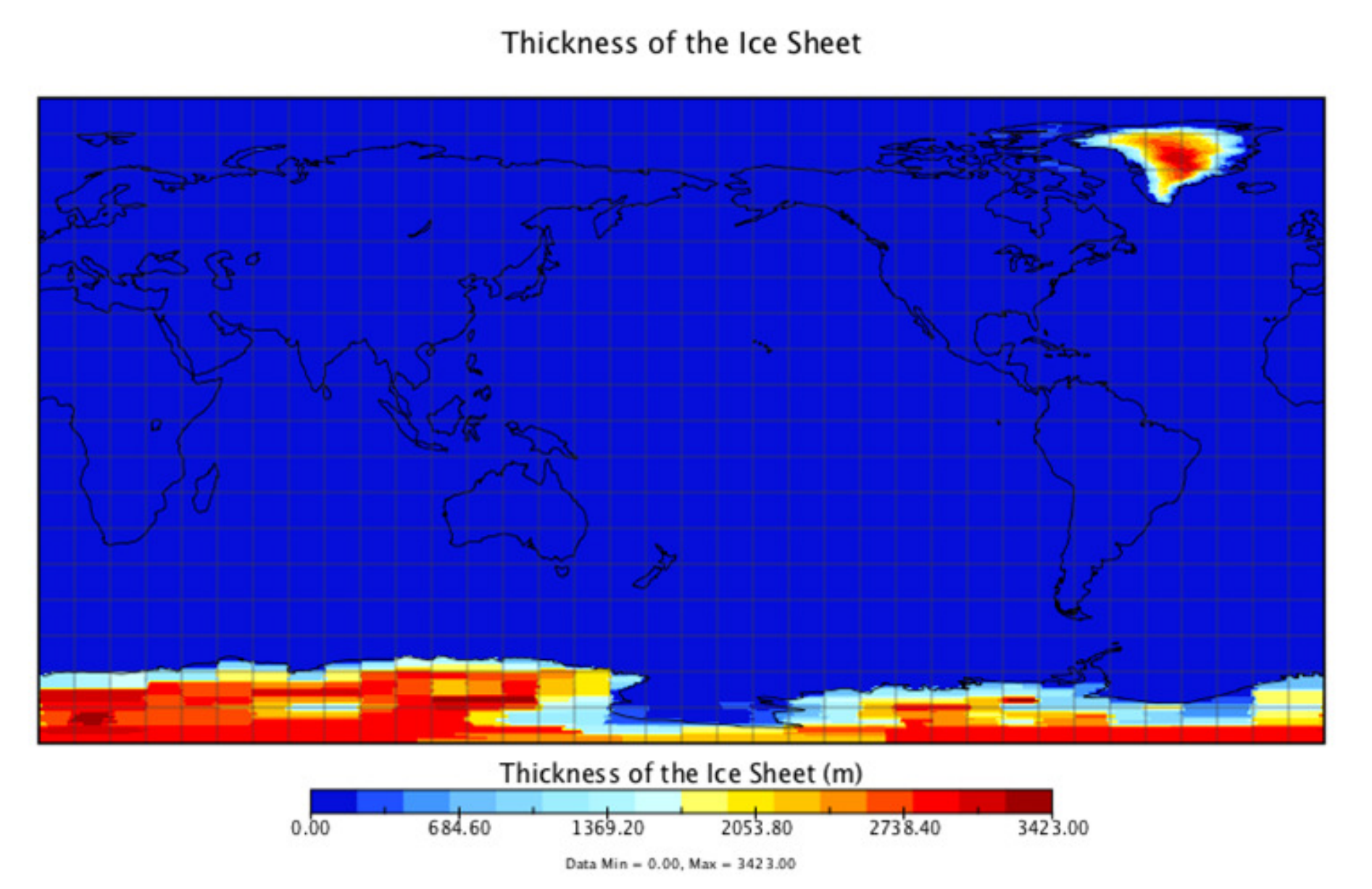}\\
  \caption{{\footnotesize
Thickness information of CRUST2.0(after \citep{Bassin-etal2000}).
 }}\label{fig-ice-coverage-thickness-corr}
\end{figure}

\begin{figure}[!ht]
   \centering
  \includegraphics[scale=0.42]{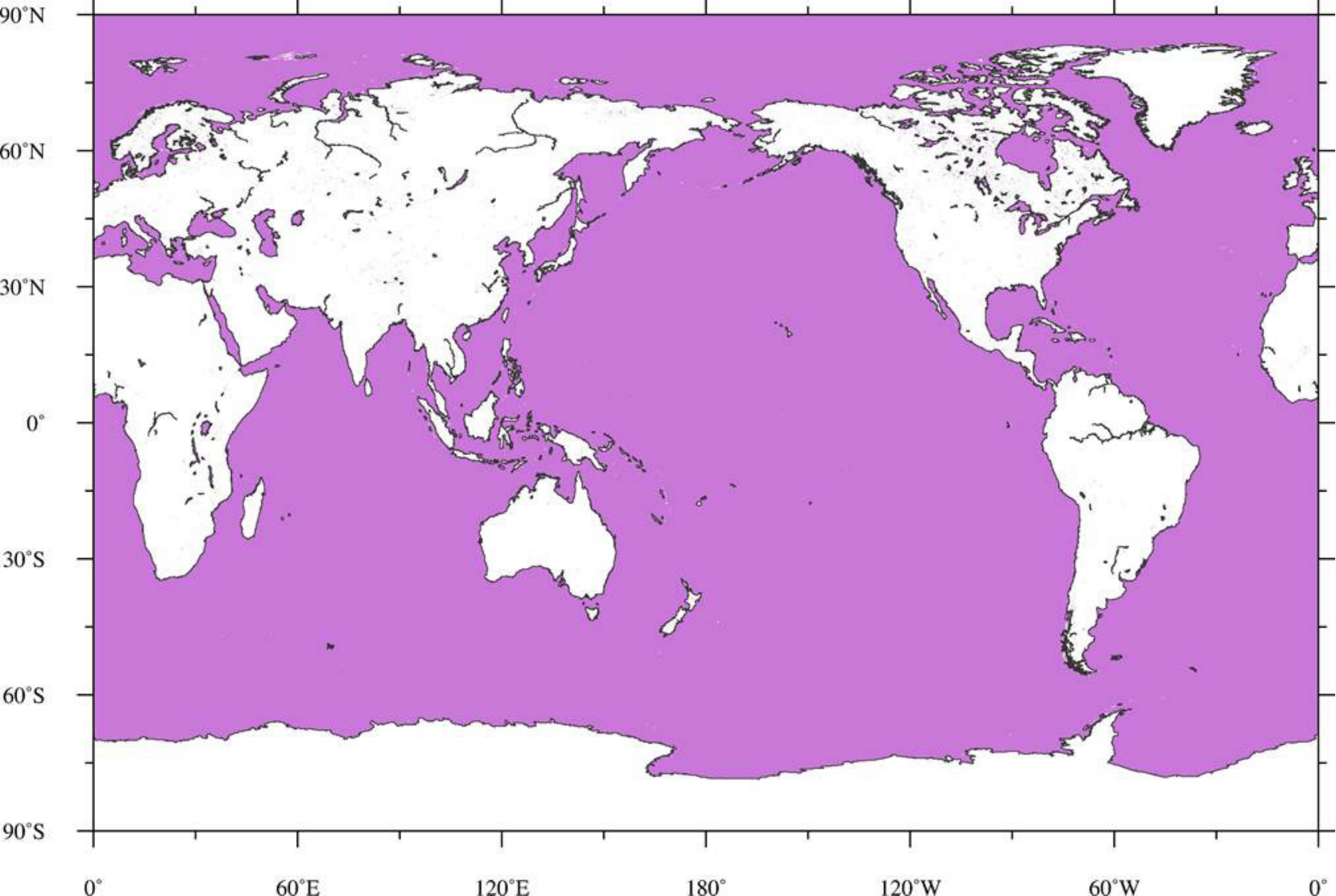}\\
  \caption{{\footnotesize
Mask (geography) derived from GMT (Generic Mapping Tools).
White color denotes the land and magenta color the ocean(after \cite{Shen-and-han2015}).
}}\label{fig-ocean-land-boundary-inf}
\end{figure}

\vspace{3mm}

\begin{figure}[!ht]
   \centering
  \includegraphics[scale=0.52]{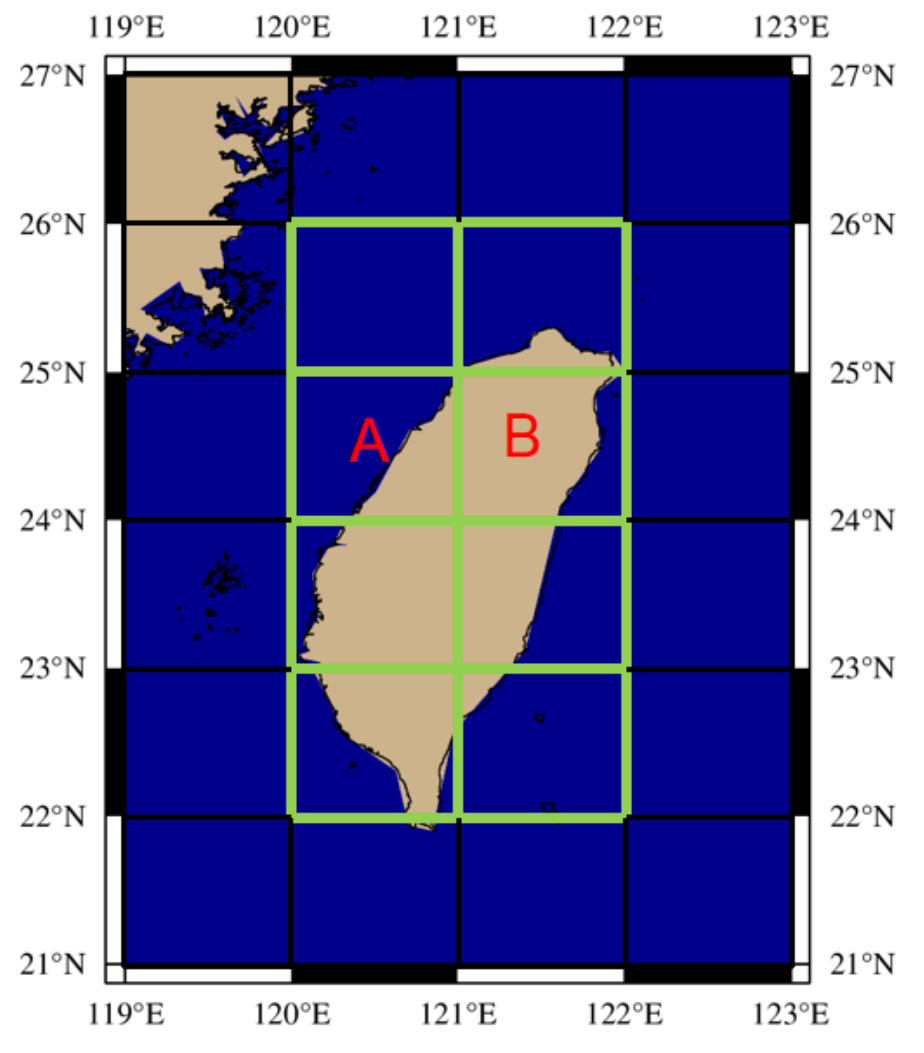}\\
  \caption{{\footnotesize
CRUST1.0 provides the following density information: Region A, with density $1.02g/cm^3$; Region B, with  density $2.10g/cm^3$(Rdrawn after \cite{Shen-and-han2015}).
 }}\label{fig-ocean-land-boundary-TW}
\end{figure}
\vspace{3mm}

\begin{figure}[!ht]
   \centering
  \includegraphics[scale=0.5]{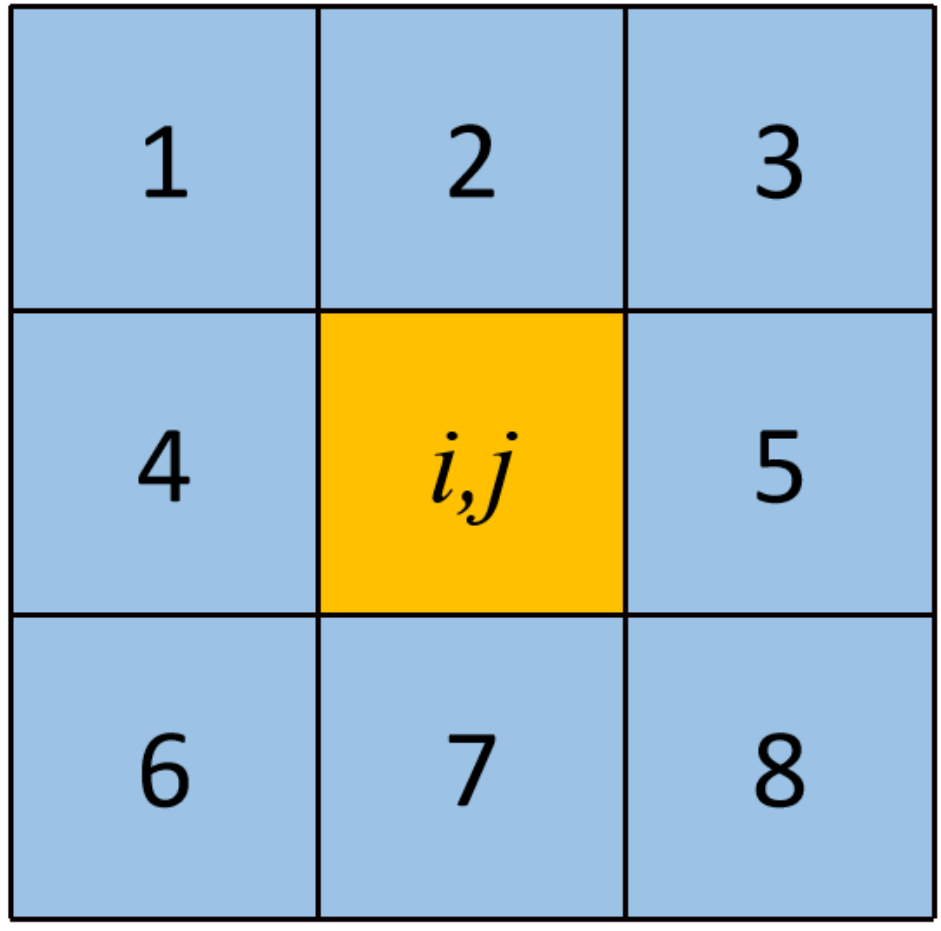}\\
  \caption{{\footnotesize
Scheme for density and thickness information corrections of $ij$-th grid. The yellow grid is $ij$-th grid and the blue grids are surrounding 8 grids(Redrawn after \cite{Shen-and-han2015}).
 }}\label{fig-grid-corr}
\end{figure}
\vspace{3mm}

\begin{figure}[!ht]
   \centering
  \includegraphics[scale=0.55]{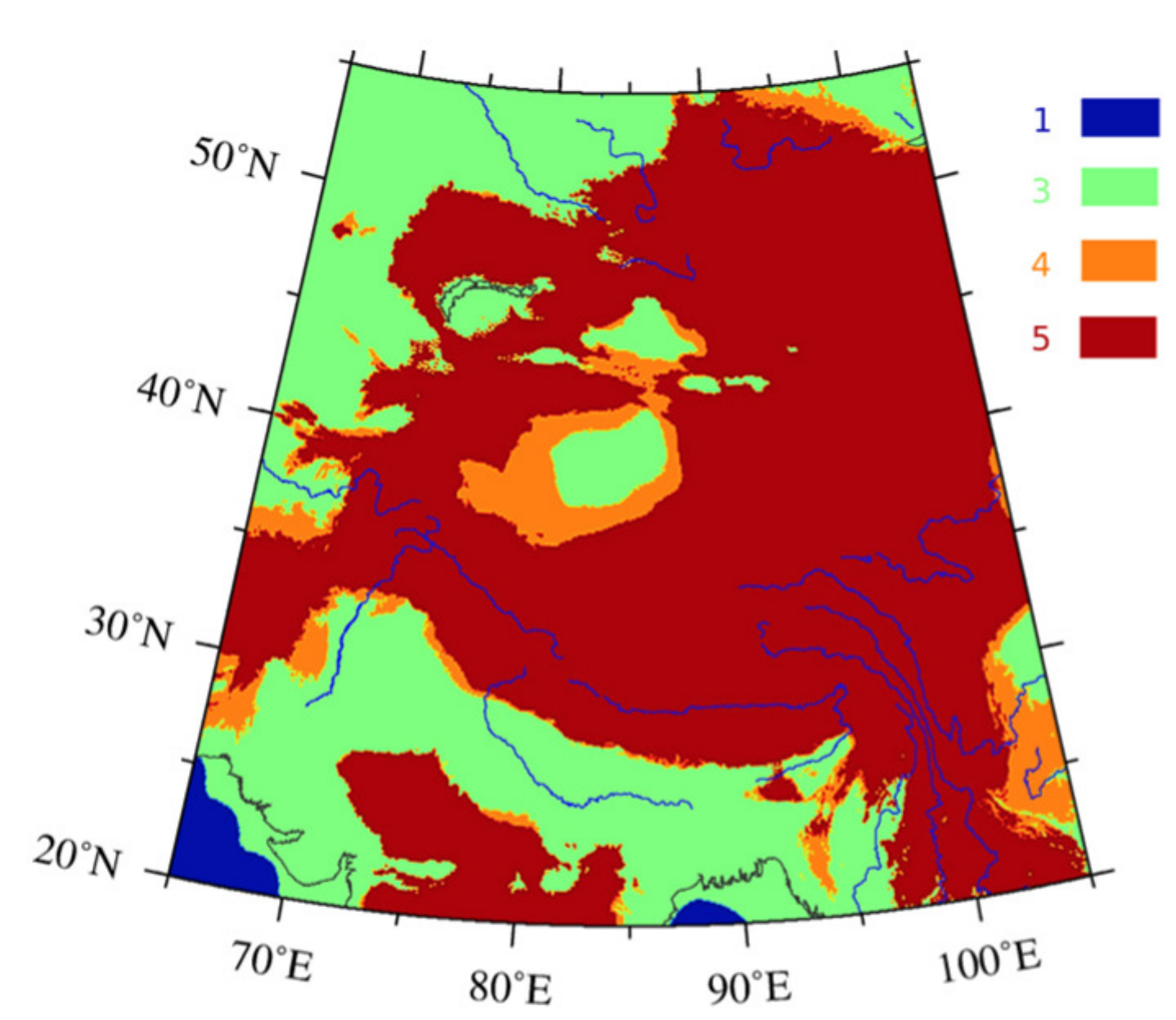}\\
  \caption{{\footnotesize
A local area case for explanation of shallow mass layer.  The information of layers and density provided by CRUST2.0$_-$re is used in the shallow mass layer model. In the area defined by this figure, only four layers of CRUST2.0$_-$re are used: (1) ice, (3) soft sediments, (4) hard sediments, and (5) upper crust (after \cite{Han2012})
 }}\label{fig-diff-layers}
\end{figure}

\vspace{3mm}

\section{Determination of the gravitational potential generated by the shallow mass layer}\label{Determination-of-the-gravitational-potential}

To determine a geoid, one needs to first determine the gravitational potential generated by the shallow mass layer. 

The gravitational potential generated by the shallow mass layer is computed by discretized numerical integration using elementary bodies such as right-rectangular prisms and tesseroids \citep{{Nagy-etal2000}, {Shen-and-Han2013}, {Tsoulis-etal2009}}. The integration of  {\color{blue} Newtonian integral} can be analytically completed by using prism modeling \citep{Nagy-etal2000,{Nagy-etal2002}} if the mass density $\rho (K)$ of each volume integral element is homogeneous. Figure \ref{fig-right-rectangular-prism} demonstrates the geometry of the right-rectangular prism. The prism is bounded by planes parallel to the coordinate planes, defined by the coordinates $X_1, X_2, Y_1, Y_2, Z_1, Z_2$, respectively, in the Cartesian coordinate system, and the field point $P$ is denoted by $(X_P, Y_P, Z_P)$.

The analytical expression of the integration is provided in the following form \citep{{Nagy-etal2000}, {Nagy-etal2002}, {Tsoulis-etal2009}, {Han-and-Shen2010}} 
\begin{eqnarray}
\label{prism-expression-eq1} 
   \begin{split}
V_1(P) & =G\rho \Bigg|\bigg|\Big| xyln\frac{z+l}{\sqrt{x^2+y^2}}+yzln\frac{x+l}{\sqrt{y^2+z^2}} \\
       &+ zxln\frac{y+l}{\sqrt{z^2+x^2}}-\frac{x^2}{2}tan^{-1}\frac{yz}{xl}\\
       &-\frac{y^2}{2}tan^{-1}\frac{zx}{yl}-\frac{z^2}{2}tan^{-1}\frac{xy}{zl}  \Big|^{x_2}_{x_1} \bigg|^{y_2}_{y_1} \Bigg|^{z_2}_{z_1} 
     \end{split}  
\end{eqnarray}
where
\begin{eqnarray}
\label{prism-expression-eq2} 
 \begin{split}
    &x_1=X_1-X_P,\hspace{3mm}x_2=X_2-X_P;\\
    &y_1=Y_1-Y_P,\hspace{3mm}y_2=Y_2-Y_P;\\
    &z_1=Z_1=Z_P,\hspace{3mm}z_1=Z_1=Z_P;\\
    &l=\sqrt{x^2+y^2+z^2}
 \end{split}
\end{eqnarray}

Eqs.(\ref{prism-expression-eq1}) and (\ref{prism-expression-eq2}) define a mathematically rigorous, closed analytical expression for the computation of the gravitational potential $V_1(P)$ of the right-rectangular prism. Although the potential  $V_1(P)$ is continuous in the entire domain $R^3$ , its solution is not defined at certain places: 8 corners, 12 edges and 6 planes of the prism \citep{Nagy-etal2000,{Nagy-etal2002}}. The direct computation of Eqs.(\ref{prism-expression-eq1}) and (\ref{prism-expression-eq2}) will fail when $P$ is located on a corner, an edge or a plane, as mentioned above, but one can calculate the corresponding limit values in a manner as suggested by \cite{Nagy-etal2000,{Nagy-etal2002}} at these special positions. Numerical calculations based on finite element approach show that Eqs.(\ref{prism-expression-eq1}) and (\ref{prism-expression-eq2}) are reliable \citep{Han-and-Shen2010}.  

The main drawback in computing the potential using Eq.(\ref{shallow-layer-integral-eq}) is the prerequisite of the repeated evaluations of several logarithmic and arctan functions for each prism. Furthermore, the formulae for computing the potential generated by prisms are given in Cartesian coordinates. This implies a planar approximation and requires a coordinate transformation for every single prism before the application of Eqs.(\ref{prism-expression-eq1}) and (\ref{prism-expression-eq2}). One needs to perform transformations between the edge system of the prism and the local vertical system at the computation point. The explicit formulae for the transformations can be found for instance in \citep{Heck-and-Seitz2007}. Due to the above reason, although the prism modeling is rigorous and exact, the corresponding computation is very time-consuming, especially when one wants to perform computations for a region with dense grid cells.

\begin{figure}[!ht]
   \centering
  \includegraphics[scale=0.55]{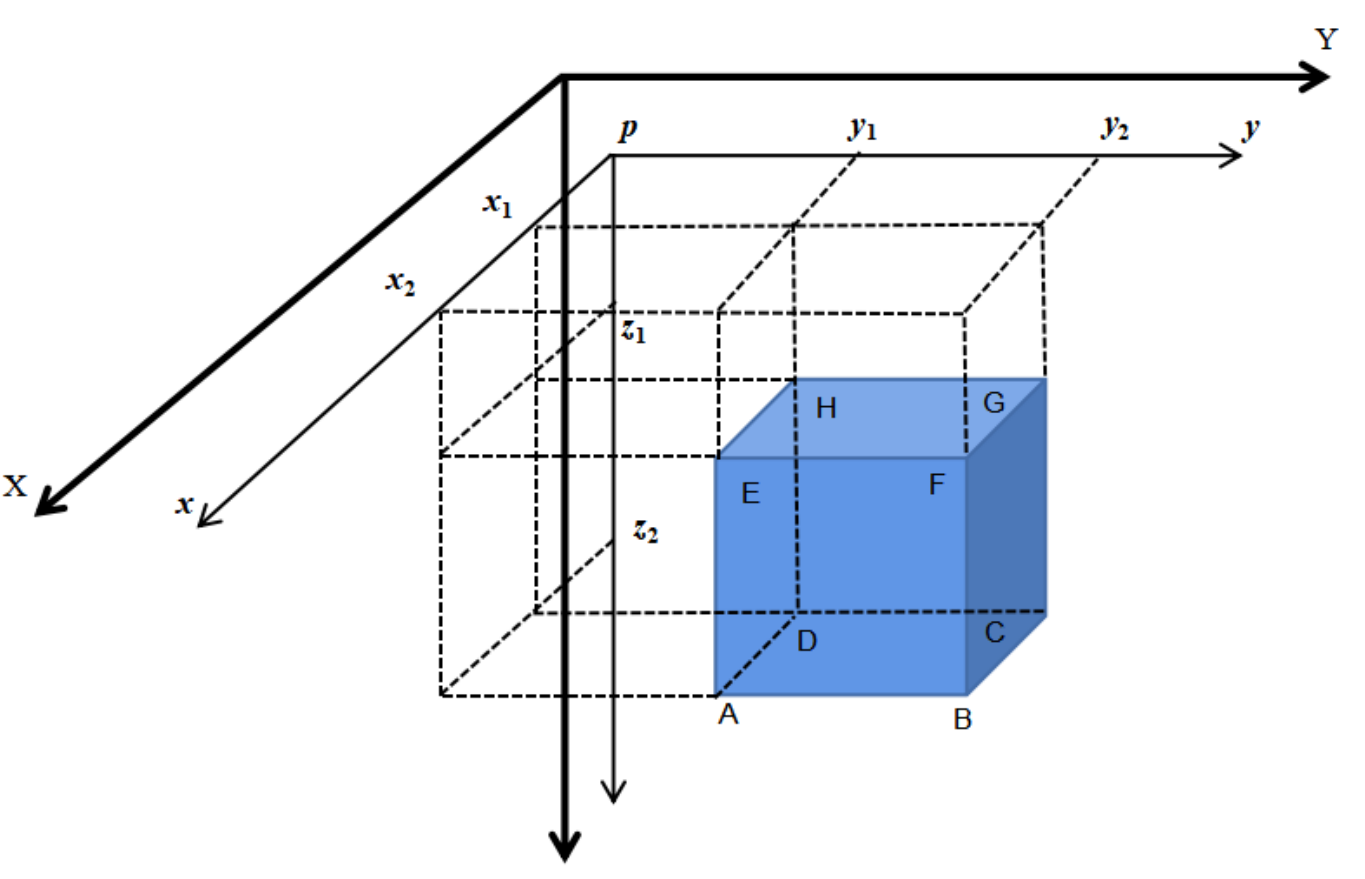}\\
  \caption{{\footnotesize
Sketch figuration of the definition of the right-rectangular prism (Redrawn after \cite{Nagy-etal2000})
 }}\label{fig-right-rectangular-prism}
\end{figure}

\vspace{3mm}

 Compared to the low efficiency of the prism modeling, the tesseroid modeling is much faster. The notion "tesseroid" (see Figure \ref{fig-tesseroid}), which was first introduced by \cite{Anderson1976}, is an elementary unit bounded by three pairs of surfaces \citep{{Kuhn2003},{Heck-and-Seitz2007},{Grombein-etal2010}, {Grombein-etal2013}}: a pair of surfaces with constant ellipsoidal heights ($r_1$=const, $r_2$=const), a pair of meridional planes ($\lambda_1$ =const, $\lambda_2$=const) and a pair of coaxial circular cones ($\varphi_1$=const, $\varphi_2$ =const). 
   
 Based on a Taylor series expansion and choosing the geometrical center of the tesseroid as the Taylor expansion point, truncated after the 3rd-order terms, the realization of Eq.(\ref{shallow-layer-integral-eq}) reads \citep{{Heck-and-Seitz2007},{Grombein-etal2013}, {Shen-and-Deng2015}} 
\begin{equation}
	\begin{split}
			V_1 (r,\varphi, \lambda) = &G\rho\Delta r\Delta \varphi \Delta \lambda [K_{000}+\frac{1}{24}(K_{200}\Delta {r^2} \\
	&+ K_{020}\Delta\varphi ^2+K_{002}\Delta \lambda ^2)+O(\Delta ^4)]
	\end{split}
	\label{second-tesseroid-expression-eq} 
\end{equation}
where $\Delta r =r_2-r_2, \Delta\varphi =\varphi_2-\varphi_1, \Delta \lambda =\lambda_2=\lambda_1$ denote the dimensions of the tesseroid,  $K_{ijk}$ denote the trigonometric coefficients involved in the Taylor expansion, the Landau symbol $O(\Delta^4)$  indicates that it contains only the 4th-order terms and higher ones, which could be neglected at present accuracy requirement. The trigonometric coefficients depend on the relative positions of the computation point ( $r,\varphi, \lambda$) with respect to the geometrical center of the tesseroid ( $r_0,\varphi_0, \lambda_0$ ). The zero-order term of Eq.(\ref{second-tesseroid-expression-eq}), which is formally equivalent to the point-mass formula, has the following form 
\begin{equation}
	\begin{split}
			& {K_{000}} = \frac{{r_o^2\cos {\varphi _o}}}{{{l_o}}},\quad {l_o} = \sqrt {{r^2} + r_o^2 - 2r{r_o}\cos {\Psi _o}} \\
& {\Psi _o} = \sin \varphi \sin {\varphi _o} + \cos \varphi \cos {\varphi _o}\cos \delta \lambda ,\quad \delta \lambda  = {\lambda _o} - \lambda 
	\end{split}
	\label{zero-tesseroid-expression-eq} 
\end{equation}

The mathematical expressions of the second-order coefficients, $K_{200}, K_{020}, K_{002}$,  are rather complicated and can be found in the work of Heck and Seitz (2007). The tesseroids are well suited to the definitions and numerical calculations of DEM/DTM, which are usually given in the form of geographical grids. The tesseroid modeling is also modest in terms of the computation costs, and more details are referred to \cite{Heck-and-Seitz2007}. 

Concerning the realization of Eq.(\ref{shallow-layer-integral-eq}), a higher order gravitational potential expression of the tesseroid modeling is referred to  \cite{Shen-and-Deng2015}.

\begin{figure}[!ht]
   \centering
  \includegraphics[scale=0.55]{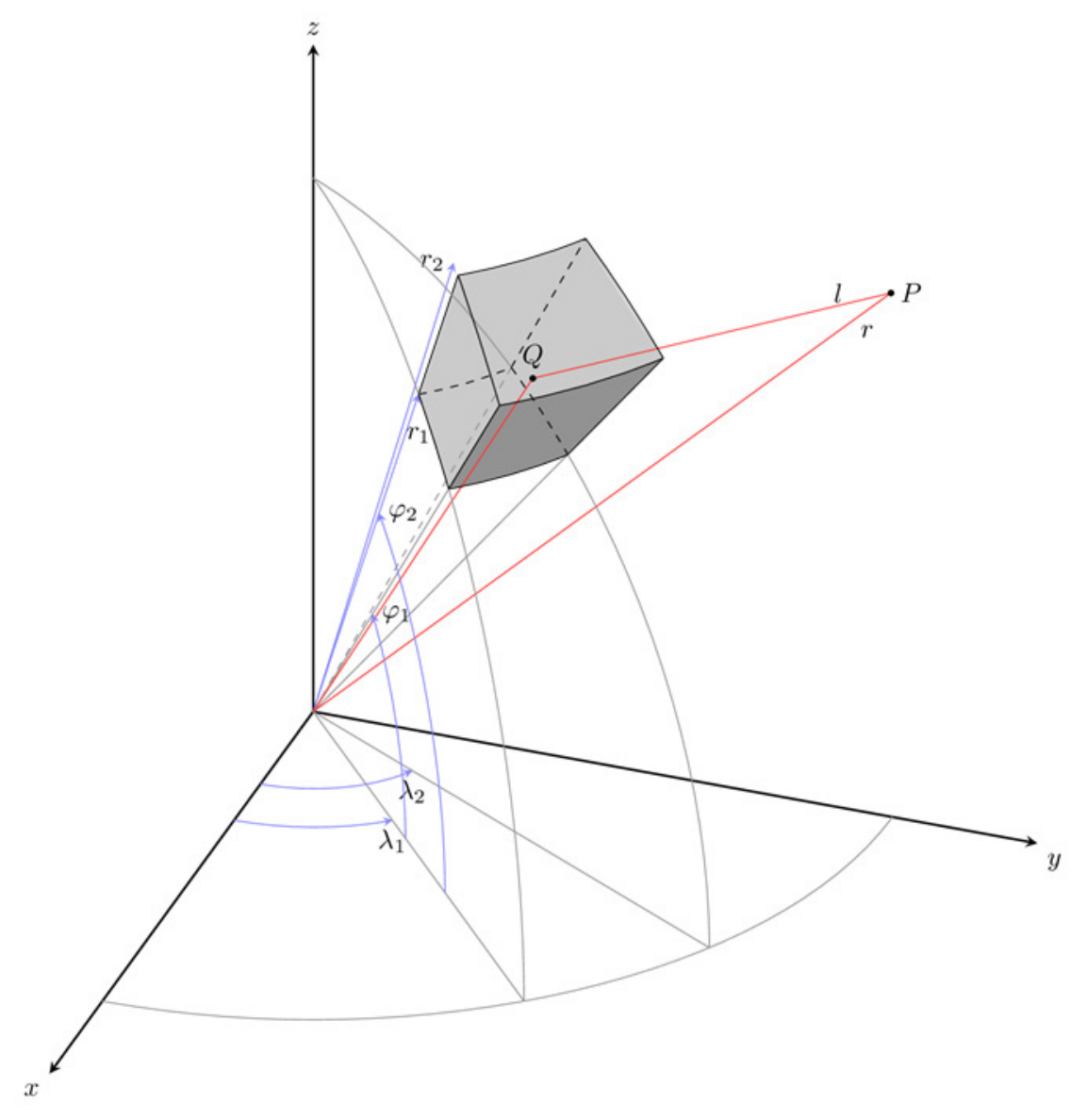}\\
  \caption{{\footnotesize
Geometry of a tesseroid (after \cite{Kuhn2003})
 }}\label{fig-tesseroid}
\end{figure}

\vspace{3mm}

The prism modeling offers relatively rigorous, analytical solution, but its implementation is inefficient and requires very demanding computations (e.g., see \cite{{Heck-and-Seitz2007},{Shen-and-Han2013}, {Shen-and-Deng2015}}). The tesseroid modeling on the other hand shows high numerical efficiency but may provide results at a relatively lower accuracy level, and the approximation errors due to the truncation of the Taylor series do exist but decrease very quickly with the increasing distance between the tesseroid and the computation point \citep{Heck-and-Seitz2007}. Hence, the ideal way is to combine these two methods together in practical computations, which would take full advantages of both methods and overcome their disadvantages at the same time \citep{Tsoulis-etal2009}. This combination method will be hereinafter referred to as the combination modeling method (CMM). In this study, we use the CMM to compute the gravitational potential of the shallow mass layer as stated in the following strategy. 

After the shallow mass layer are partitioned into elementary units, the prism modeling is adopted to precisely calculate the contribution of the units which are located at the nearest vicinity surrounding the computation point, while the tesseroid modeling is employed for computing the contribution of the units located outside the mentioned vicinity area (see details in Section \ref{Global geoid model  2022 (GGM2022)}). In this case, one can maintain a reasonable computation cost  without losing the required accuracy. Our previous experimental studies show that the differences between prism method and the CMM are quite small, with an order of $\sim 10^{-3}$ m$^2$s$^2$. That means the corresponding difference in height is less than 1mm.

\section{Global geoid model  2022 (GGM2022)}\label{Global geoid model  2022 (GGM2022)}

The integration of the gravitational potential of the shallow mass layer actually contains three zones: $1^\circ$ spherical near-zone, $1^\circ$ to $10^\circ$ spherical translation zone, and the far-zone. Owing to the high computation cost, the efficiency of the entire computation requires an appropriate choice of the modeling methods and computation strategies. The CMM is used as described in Section \ref{Determination-of-the-gravitational-potential}.

To keep the accuracy and raise efficiency, in different zones we use different formulas: in $1^\circ$ spherical near-zone, we use prism formula , namely, Eqs.(\ref{prism-expression-eq1}) and (\ref{prism-expression-eq2}), with a $5^\prime \times 5^\prime$ DEM; in $1^\circ$ to $10^\circ$ spherical translation zone, we use second-order tesseroid formula(\cite{Nagy-etal2000}), namely, Eq.(\ref{second-tesseroid-expression-eq}), with a $15^\prime \times 15^\prime$ DEM; and in the far-zone, we use the zero-order tesseroid formula, namely, Eq.(\ref{zero-tesseroid-expression-eq}), with a $30^\prime \times 30^\prime$ DEM. In the ocean area, we use DNSC08 mean sea surface model. 

The refined $5^\prime \times 5^\prime$ global crustal model CRUST$_-$re (both density and stratification information) is used in the near-zone,  de-sampled to a $15^\prime \times 15^\prime$ grid in the translation zone and to a $30^\prime \times 30^\prime$  grid in the far-zone, respectively, corresponding to grid sizes in these three zones. 

The gravitational effects for all zones are computed on a spatial sphere enclosing the whole Earth, with a radius $R_B =6386$ km from the center of the Earth (noted as to the Brillourin radius). It is noted that the semi-major axis of the reference ellipsoid is $\sim 6378$ km, and the highest mountain in the world is $\sim 8$ km at a latitude around $27^\circ$N, so a sphere with a radius $R_B = 6386$ km can include all the masses of the Earth. The gravitational potentials generated from the shallow mass layer is  shown in Figure \ref{fig-shallow-layer-potential}.  And the  static results of shallow layer potential are shown in table\ref{tab- shallow-layer-potential}

\begin{figure}[!ht]
   \centering
\leftline{(a)}
  \includegraphics[scale=0.44]{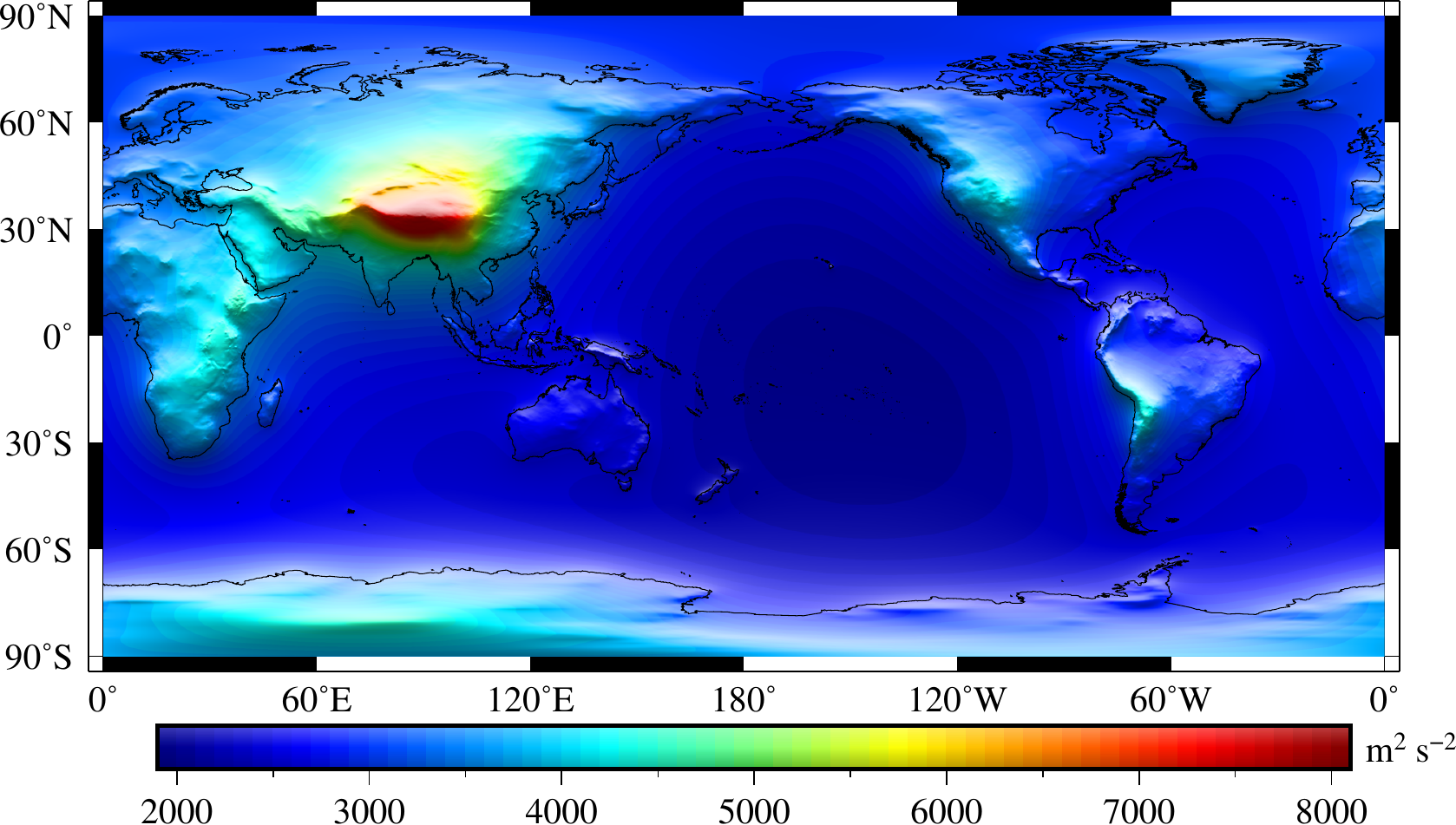}\\

  \vspace{2mm}
\leftline{(b)}
  \includegraphics[scale=0.44]{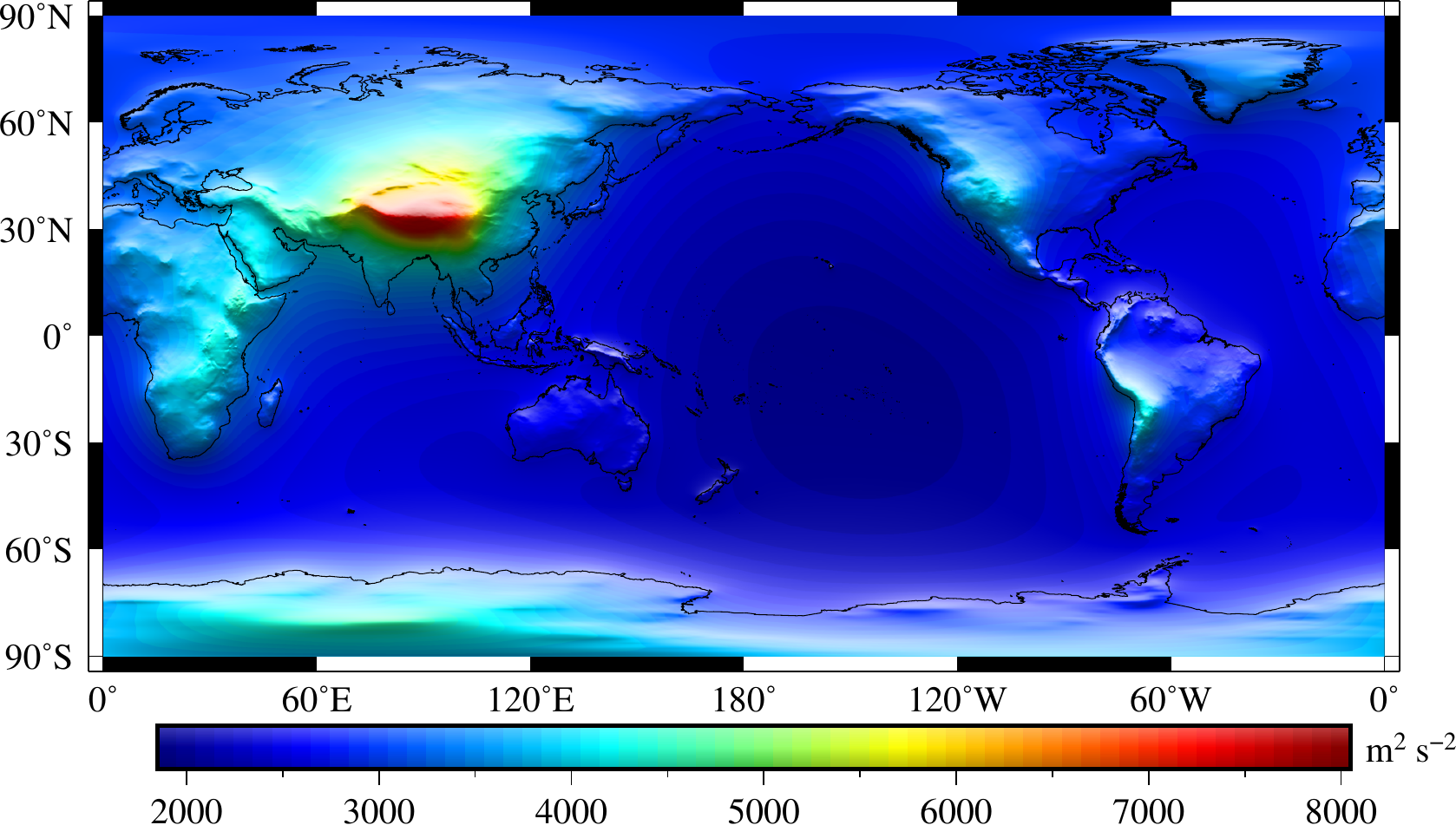} \\

  \caption{{\footnotesize
Gravitational potential $V_1(P)$ generated by the shallow mass layer, on the spherical surface with its radius $R=6386$ km. (a) top slot  calculated by CRUST1.0,  and  (b)  bottom slot calculated  by CRUST2.0.
 }}\label{fig-shallow-layer-potential}
\end{figure}

%statistics of shallow-layer-potential
\begin{table}
   \centering
 \caption{The statistics information of shallow-layer-potential on the spherical surface with its radius $R=6386$ km (Unit: m$^2$/s$^2$)}
\label{tab- shallow-layer-potential}
\begin{tabular}{ccccc}
\toprule
Model &  Min & Max &  Mean & Std   \\
\midrule
Crust1.0 & 1908.82 & 8027.96 &  2870.55 &740.79 \\
Crust2.0 & 1866.29 & 7851.07 & 2820.28 & 729.34 \\
\bottomrule
\end{tabular}
\end{table}

The spherical harmonic approach is used to determine the gravitational potential $V^*_0(P)$  in the domain $\Omega_\Gamma ^c$ , and the procedures are described as follows.

(i) Expand the gravitational effects of the shallow mass layer up to degree and order 2190 in spherical harmonics, and the derived coefficients are denoted as (${\bar C}^1_{nm},{\bar S}^1_{nm}$). In order to link the expansion coefficients with those of EGM2008, the same scale factor, namely, the geocentric gravitational constant $GM$, is used in the spherical harmonic expansion, which is expressed as \citep{Shen-and-Han2013}

\begin{equation}
\begin{split}
\left\{ {\begin{array}{*{20}{c}}
{\bar C_{nm}^1}\\
{\bar S_{nm}^1}
\end{array}} \right\} =& \frac{{{R_B}}}{{4\pi GM}} \times \sum\limits_{i = 1}^I {\sum\limits_{j = 1}^J {{V_1}} } ({\theta _i},{\mkern 1mu} {\lambda _j})\times \\
& {{\bar P}_{nm}}(\cos {\theta _i})  
	\left\{ {\begin{array}{*{20}{c}}
{\cos m{\lambda _j}}\\
{\sin m{\lambda _j}}
\end{array}} \right\}{w_i}
\end{split}
	\label{cmg7-2-5} 
\end{equation}
where 
\begin{equation}
\label{cmg7-2-6} 
{w_i} = {R_S}^2\frac{{4\pi }}{{{n^2}}}\sin {\theta _i}\sum\limits_{l = 0}^{n/2 - 1} {\frac{{\sin ((2l + 1){\theta _i})}}{{2l + 1}}} n,
\end{equation}
$n/m$ is the degree/order used in the spherical harmonic analysis, and the truncated degree is 2190;  $({\theta _i},\,\,{\lambda _j})$ are co-latitude and longitude of the geometric center of the $ij$-th grid, $V({\theta _i},\,\,{\lambda _j})$ is the gravitational potential of the shallow mass layer on the spatial sphere, $w_i$ are the area weights given by \cite{Driscoll-and-Healy1994};  ${\bar P}_{nm}$ denote the fully normalized associated Legendre functions of degree $n$ and order $m$. For the sake of numerical stability, the function ${\bar P}_{nm}$  with high degree and order are calculated using the recursive algorithms as described by \cite{Holmes-and-Featherstone2002} to avoid the underflow and overflow.

(ii) Use a simple yet effective quality control of the spherical harmonic expansion to perform the reverse process of (i), and a spherical harmonic synthesis is used to compute $V^{(1)}_1$ from the spherical harmonics ($\bar C_{nm}^1, \bar S_{nm}^1$), expressed as 
\begin{equation}
\label{V1(1)-harmonic-eq} 
\begin{split}
V_1^{(1)}(\theta , \lambda) = & \frac{GM}{R_B}[1 + \sum\limits_{n = 2}^M (\frac{a}{R_B})^n \sum\limits_{m = 0}^n (\bar C_{nm}^1\cos m\lambda  \\
& + \bar S_{nm}^1\sin m\lambda ){\bar P}_{nm}(\cos \theta )]
\end{split} 
\end{equation}
And then the following residual 
\begin{equation}
\label{DeltaV1(1)-eq} 
\Delta V_1^{(1)} = {V_1} - V_1^{(1)}
\end{equation}
is checked to see whether the criterion a priori given is fulfilled, which could be set according to the accuracy requirement.  

(iii) Perform the spherical harmonic analysis again
\begin{equation}
\begin{split}
\left\{ {\begin{array}{*{20}{c}}
{\Delta \bar C_{nm}^1}\\
{\Delta \bar S_{nm}^1}
\end{array}} \right\} =& \frac{{{R_B}}}{{4\pi GM}} \times \sum\limits_{i = 1}^I {\sum\limits_{j = 1}^J {{\Delta V_1}} } ({\theta _i},{\mkern 1mu} {\lambda _j})\times \\
& {{\bar P}_{nm}}(\cos {\theta _i})  
	\left\{ {\begin{array}{*{20}{c}}
{\cos m{\lambda _j}}\\
{\sin m{\lambda _j}}
\end{array}} \right\}{w_i}
\end{split}
	\label{Delat-CS1nm-harmonic-eq} 
\end{equation}
and update ($\bar C_{nm}^1, \bar S_{nm}^1$) with the following relations
\begin{equation}
\begin{array}{*{20}{c}}
{\bar C_{nm}^{1(1)} = \bar C_{nm}^1 + \Delta \bar C_{nm}^1}\\
{\bar S_{nm}^{1(1)} = \bar S_{nm}^1 + \Delta \bar S_{nm}^1}
\end{array}
	\label{Delta-CS1(1)-eq} 
\end{equation}

(iv) Repeat steps (ii) and (iii) until the iteration converges, namely after $N$ steps one obtains ($\bar C_{nm}^{1(N)}, \bar S_{nm}^{1(N)}$), and the standard deviation (STD) of $\Delta V^{(N)}_1$  does not change significantly. The basic concern behind this iteration process lies in the fact that the results provided by only one spherical harmonic analysis step are not accurate enough. $N$ depends on the accuracy required. For instance, in the study of \cite{Shen-and-Han2013}, the results converge after three iterations ($N=3$). The residuals $\Delta V^{(3)}_1$ range from $-1.7$ to $1.1$ m$^2$/s$^2$, with a mean of almost zero and STD of $0.16$ m$^2$/s$^2$, which is equivalent to $1.6$ cm in height. If a higher accuracy is needed, we need more iterations. Relevant details are referred to \cite{Shen-and-Han2013}.
And in this paper, the results converge after four hundreds iterations ($N=400$). Because we use two kinds of CRUST models, so the static results are shown in the table \ref{tab- iterations}.

\begin{table}
   \centering
 \caption{The statistics information of residuals of Spherical harmonic coefficients for shallow layer after four hundreds iterations (unit: m).  } 
\label{tab- iterations}
\begin{tabular}{ccccc}
\toprule
Model &  Min & Max &  Mean & Std   \\
\midrule
Crust1.0 & -6.11E-02 & 6.16E-02 & 5.76E-014 & 4.82E-04 \\
Crust2.0 & -5.87E-02 & 6.20E-02 &-1.35E-013 & 4.72E-04 \\
\bottomrule
\end{tabular}
\end{table}

\vspace{3mm}

(v) Subtract the coefficients ($\bar C_{nm}^{1(N)}, \bar S_{nm}^{1(N)}$) from the EGM2008 spherical harmonics ($\bar C_{nm}, \bar S_{nm}$) in order to get the spherical harmonic representation ($\bar C_{nm}^{0*}, \bar S_{nm}^{0*}$) of the masses enclosed by the surface $S$ (inner mass)

\begin{equation}
\begin{array}{l}
\bar C_{nm}^{0*} = {{\bar C}_{nm}} - \bar C_{nm}^1\\
\bar S_{nm}^{0*} = {{\bar S}_{nm}} - \bar S_{nm}^1
\end{array}
	\label{CSnm(0)-eq} 
\end{equation}

(vi) Compute the gravitational potential generated by the inner mass, $V_0^*(P) (P \in \bar S )$ from ($\bar C_{nm}^{*}, \bar S_{nm}^{*}$), expressed as

\begin{equation}
\label{V0-star-harmonic-eq} 
\begin{split}
V_0^{*}(\theta , \lambda,r) = & \frac{GM}{r}[1 + \sum\limits_{n = 2}^M (\frac{a}{r})^n \sum\limits_{m = 0}^n (\bar C_{nm}^{0*}\cos m\lambda  \\
& + \bar S_{nm}^{0*}\sin m\lambda ){\bar P}_{nm}(\cos \theta )]
\end{split} 
\end{equation}
where $r$ is the distance between the field point and the geometric center of the Earth. The geopotential $W(P)$ in domain $\bar{S}$  can be determined once $V^*_0$ is determined. 

In order to determine the geoid based on Eq.(\ref{geoid-eq}), an iterative procedure as described in Section 2.2 can be applied.

%\begin{figure}[!h]
%   \centering
 % \includegraphics[scale=0.45]{fig-shallow-layer-potential-diff-eps-converted-to.pdf}\\
 % \caption{{\footnotesize
%Differences between original potential $V_1(P)$ and the corresponding harmonic synthesis result $V_{1*}(P)$ (unit: m$^2$/s$^2$)
% }}\label{fig-shallow-layer-potential-diff}
%\end{figure}

%\vspace{3mm}

\section{Evaluation of the GGM2022}
\label{Evaluation of the GGM2022}

  Solution of geoid equation (\ref{geoid-eq}) provides a $5^\prime \times 5^\prime$ global geoid model (GGM2022), which is shown by Figure \ref{GGM2020}. And the statistics values are shown by table  \ref{tab- GGM2020}

%\section{Evaluation of GGM2015}
In the sequel we provide two kinds of evaluations. First, GGM2022 is compared to EGM2008 global geoid (E08GG). Second, we simultaneously compare GGM2022 and E08GG with globally available GPS leveling points. 

\begin{figure}[!ht]
   \centering
\leftline{(a)}
  \includegraphics[scale=0.44]{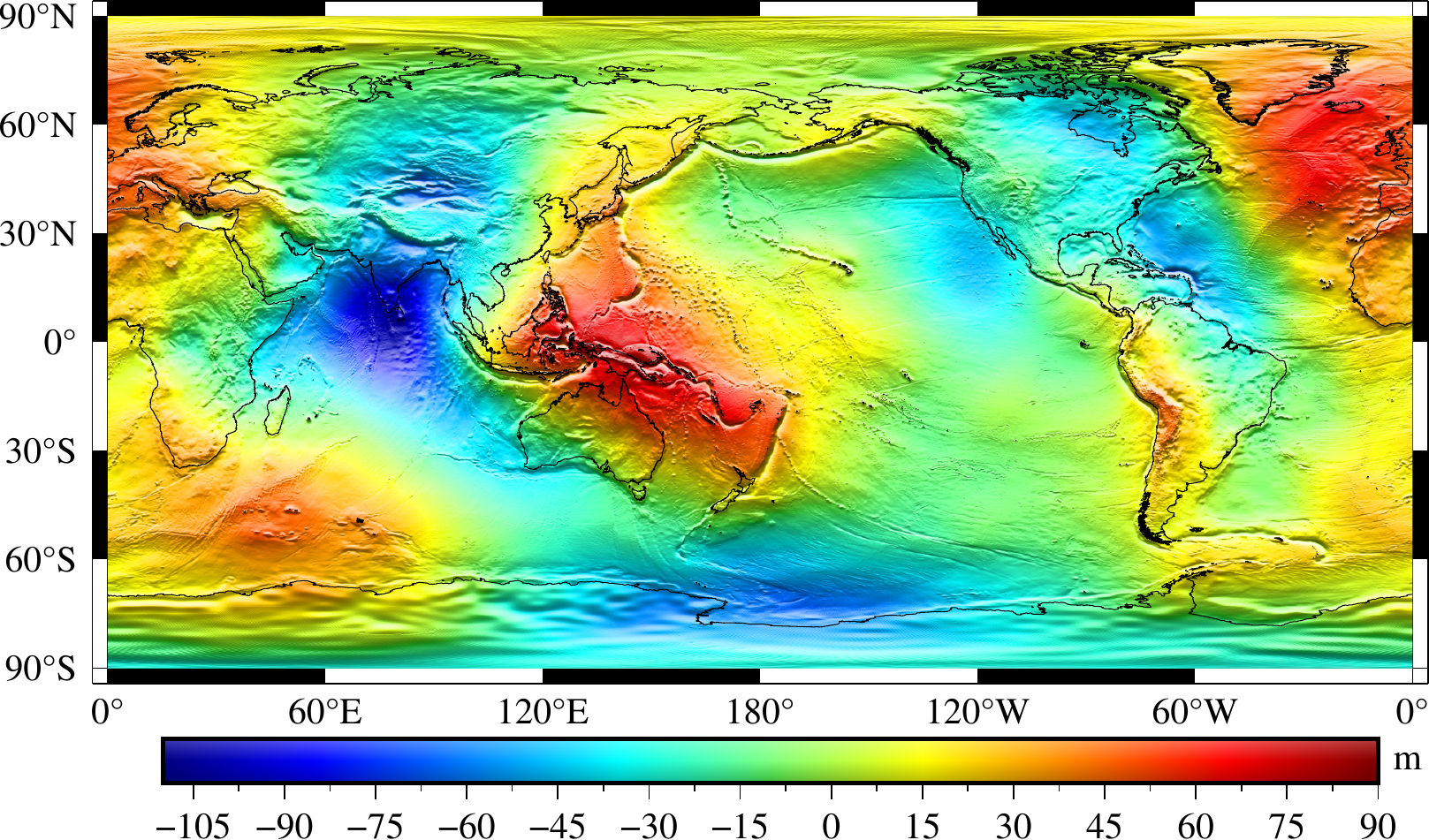}\\

  \vspace{2mm}
\leftline{(b)}
  \includegraphics[scale=0.44]{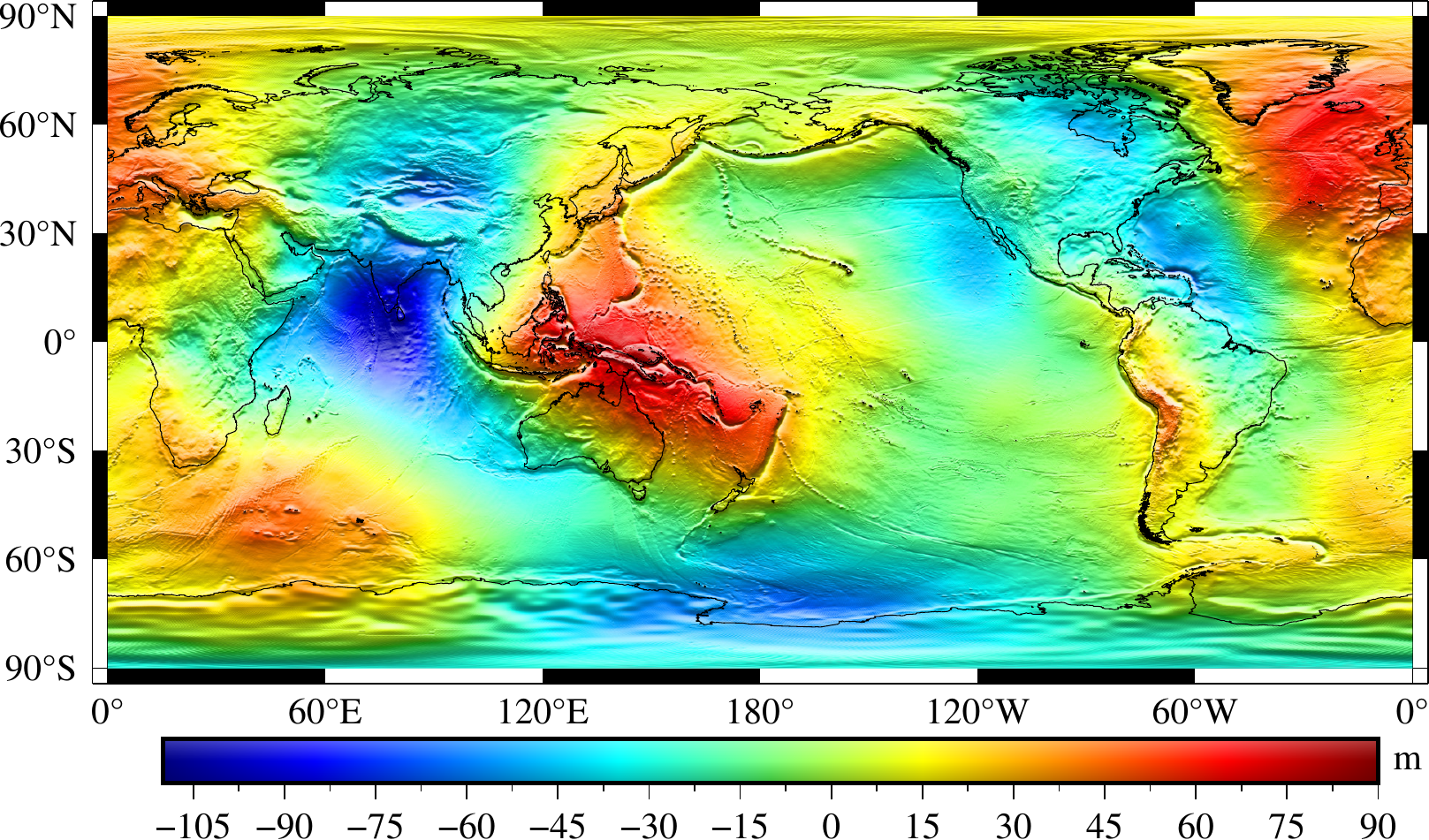}\\
  
  \caption{{\footnotesize
The calculated $5^\prime \times 5^\prime$ global geoid model 2022 (GGM2022).   (a) top slot calculated by CRUST1.0,  and (b)  bottom slot calculated  by CRUST2.0.
 }}\label{GGM2022}
\end{figure}
%\vspace{3mm}

\begin{table}
   \centering
 \caption{The statistics information of  global geoid model 2022 (GGM2022) calculated by CRUST models, compared with the EGM2008 geoid model (Unit: m). }
\label{tab- GGM2022}
\begin{tabular}{ccccc}
\toprule
Model &  Min & Max &  Mean & Std   \\
\midrule
  GGM22$_-1.0$ $^*$ & -106.50 & 86.52 & -0.87& 29.13 \\
 GGM22$_-2.0$ $^*$   & -106.50 & 86.62 &-0.87 & 29.13 \\
   E08GG$^*$ & -106.51 & 86.29 &-0.91 & 29.06 \\
\bottomrule
\end{tabular}\\
  {\tiny $^*$ GGM22$_-1.0$$^*$ and GGM22$_-2.0$ $^*$ stand for GGM2022 based on CRUST1.0 and  CRUST2.0, respectively; E08GG$^*$ stands for EGM2008 Global Geoid}
\end{table}
\vspace{3mm}

\subsection{Comparison with the EGM2008 global geoid}\label{Comparison-with-the-EGM2008-global-geoid-subsection}

The EGM2008 geopotential model \citep{Pavlis-etal2008} and EGM2008 global geoid (E08GG) play an important role in describing the geopotential field generated by the Earth in the domain $\Omega_S ^c$ and world height datum system. Here, we use the EGM2008 geopotential model, but we calculate the geoid in a manner quite different from the conventional one (e.g., \cite{Pavlis-etal2008}). Then, it is necessary to demonstrate the differences between the calculated global geoid model 2022 (GGM2022) and E08GG. We evaluate the differences between the two geoid models over the globe where the GPS leveling points are available. The GGM2022 is shown in Figure \ref{GGM2022}, with the differences between these two geoid models shown in Figure \ref{fig-Diff-GGM2020-and-E08GG}. The statistics results of their comparisons are listed in Table \ref{tab-Diff-GGM2020-and-E08GG}.  Significant differences between them are found in rough mountainous areas (e.g., Tianshan Mountain, Kunlun Mountain and Himalaya Mountain), and polar areas (Antarctic and Arctic areas) as shown by Figures\, \ref{fig-Antarctic-Diff-GGM2020-and-E08GG} and \ref{fig-Arctic-Diff-GGM2020-and-E08GG}, respectively. In ocean regions, there are little difference between E08GG and GGM2022. To evaluate which model is better, we need to compare both E08GG with GGM2022 with GPS leveling observations.

\begin{figure}[!ht]
   \centering
\leftline{(a)}
  \includegraphics[scale=0.44]{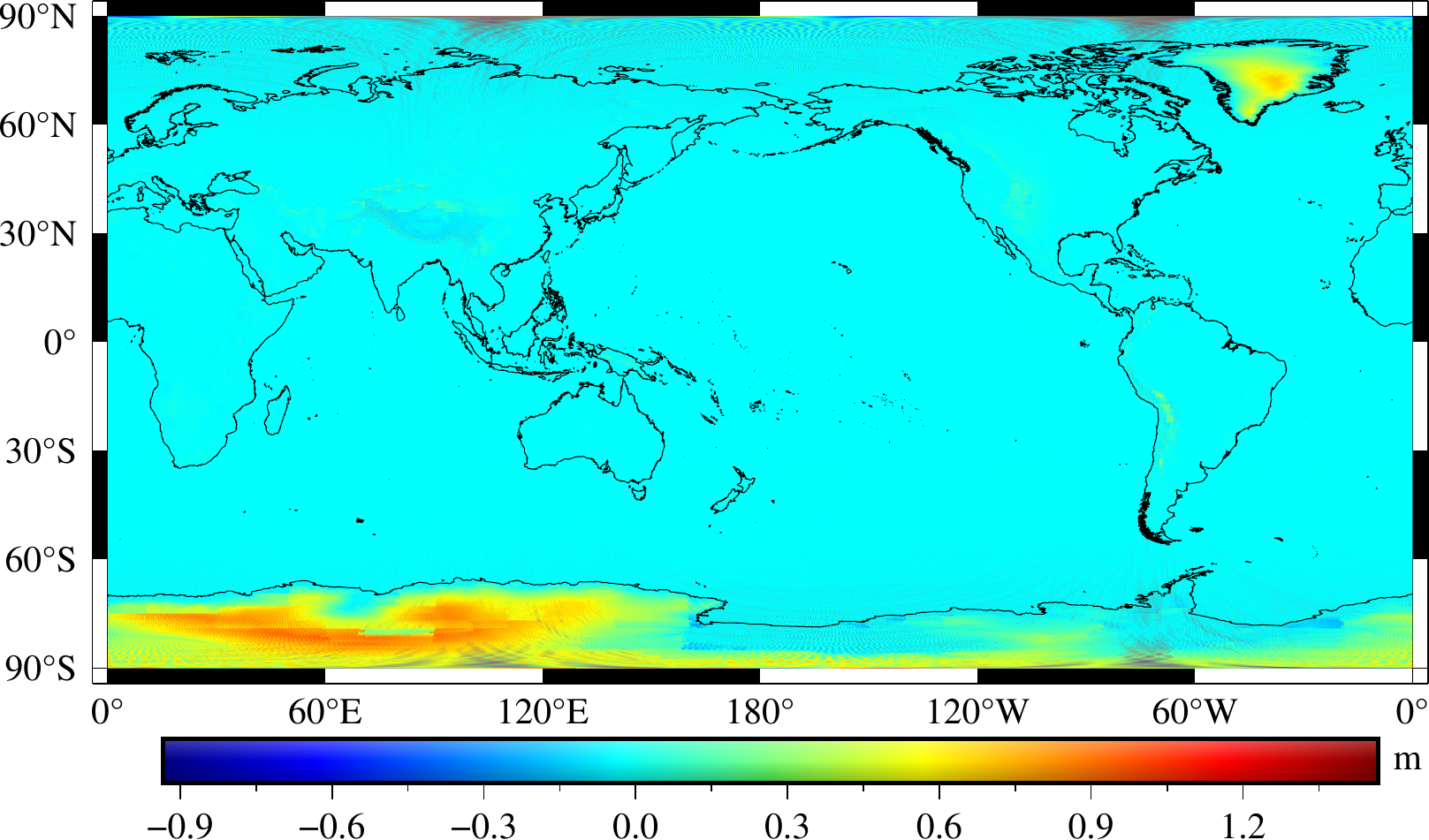}\\

  \vspace{2mm}
\leftline{(b)}  
  \includegraphics[scale=0.44]{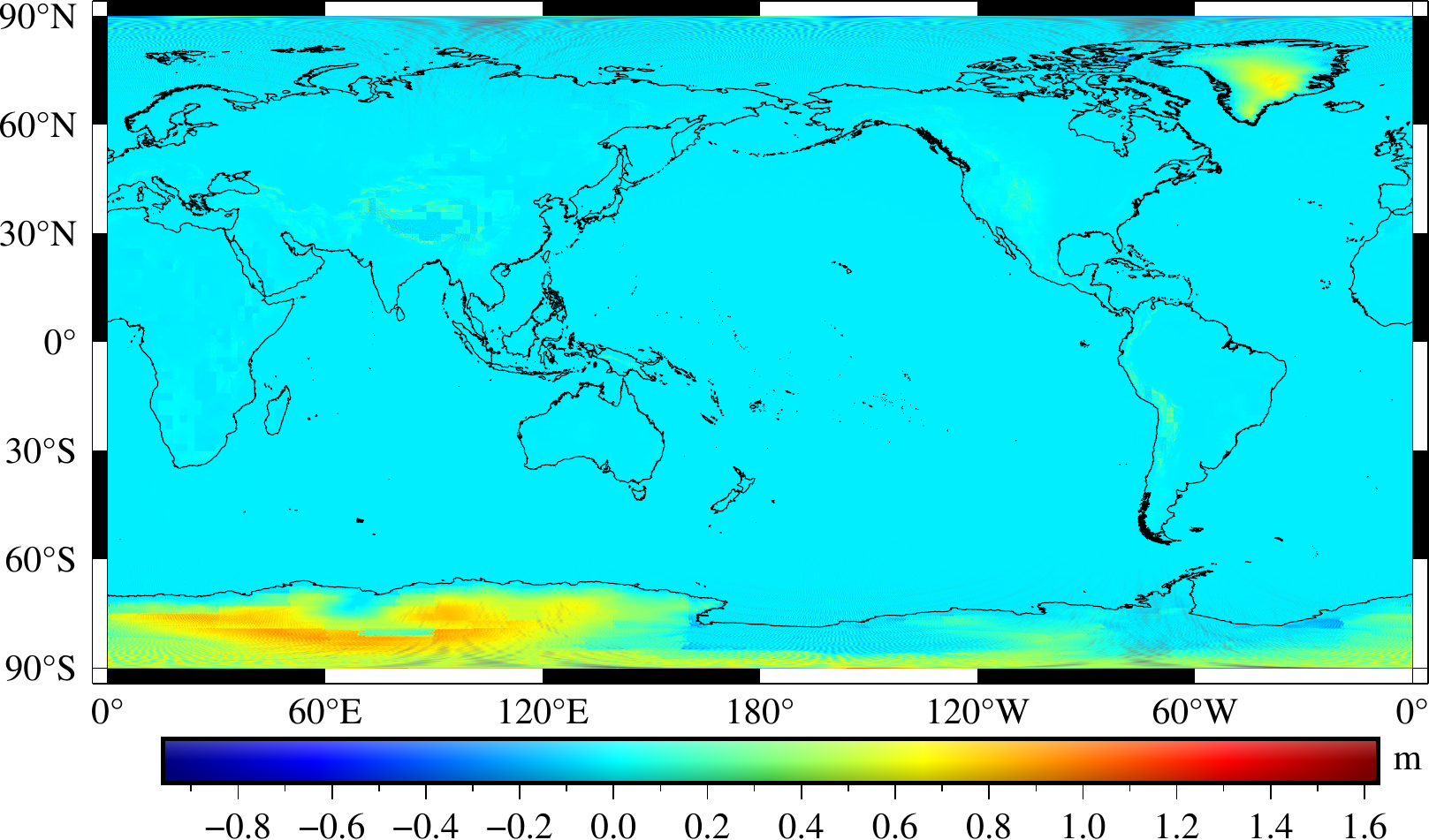}\\

  \caption{{\footnotesize
Differences between GGM2022 and EGM2008 global geoid (E08GG), (a) top slot  calculated by CRUST1.0, and (b)  bottom slot calculated  by CRUST2.0 (unit: m) 
 }}\label{fig-Diff-GGM2020-and-E08GG}
\end{figure}

\vspace{3mm}

\begin{table}
   \centering
 \caption{The statistics information of differences between GGM2022 and EGM2008 global geoid (Unit: m)  }
\label{tab-Diff-GGM2020-and-E08GG}
\begin{tabular}{ccccc}
\toprule
Model &  Min & Max &  Mean & Std   \\
\midrule
  GGM22$_-$1.0$^*$ -  E08GG$^*$  & -0.93 & 1.46 & 5.69E-02 &0.15 \\
  GGM22$_-$2.0$^*$ -  E08GG$^*$ & -0.96 & 1.62 & 6.18E-02 &0.15 \\
\bottomrule
\end{tabular}
 {\tiny $^*$ GGM22$_-$1.0$^*$ and GGM22$_-$2.0$^*$ stand for GGM2022 based on CRUST1.0 and  CRUST2.0, respectively; E08GG$^*$ stands for EGM2008 Global Geoid}
\end{table}

\begin{figure}[!ht]
   \centering
  \includegraphics[scale=0.55]{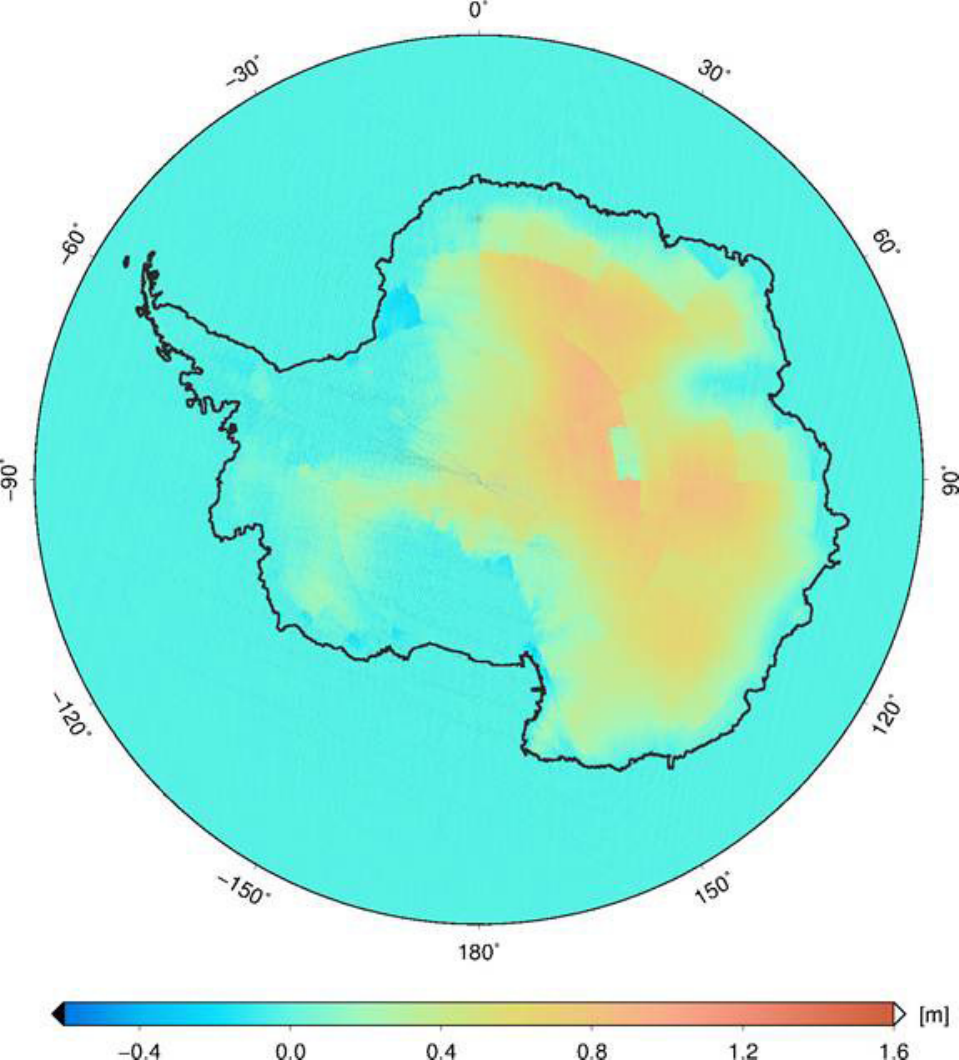}\\
  \caption{{\footnotesize
Differences between GGM2022 and E08GG in Antarctic area
 }}\label{fig-Antarctic-Diff-GGM2020-and-E08GG}
\end{figure}

\vspace{3mm}

\begin{figure}[!ht]
   \centering
  \includegraphics[scale=0.55]{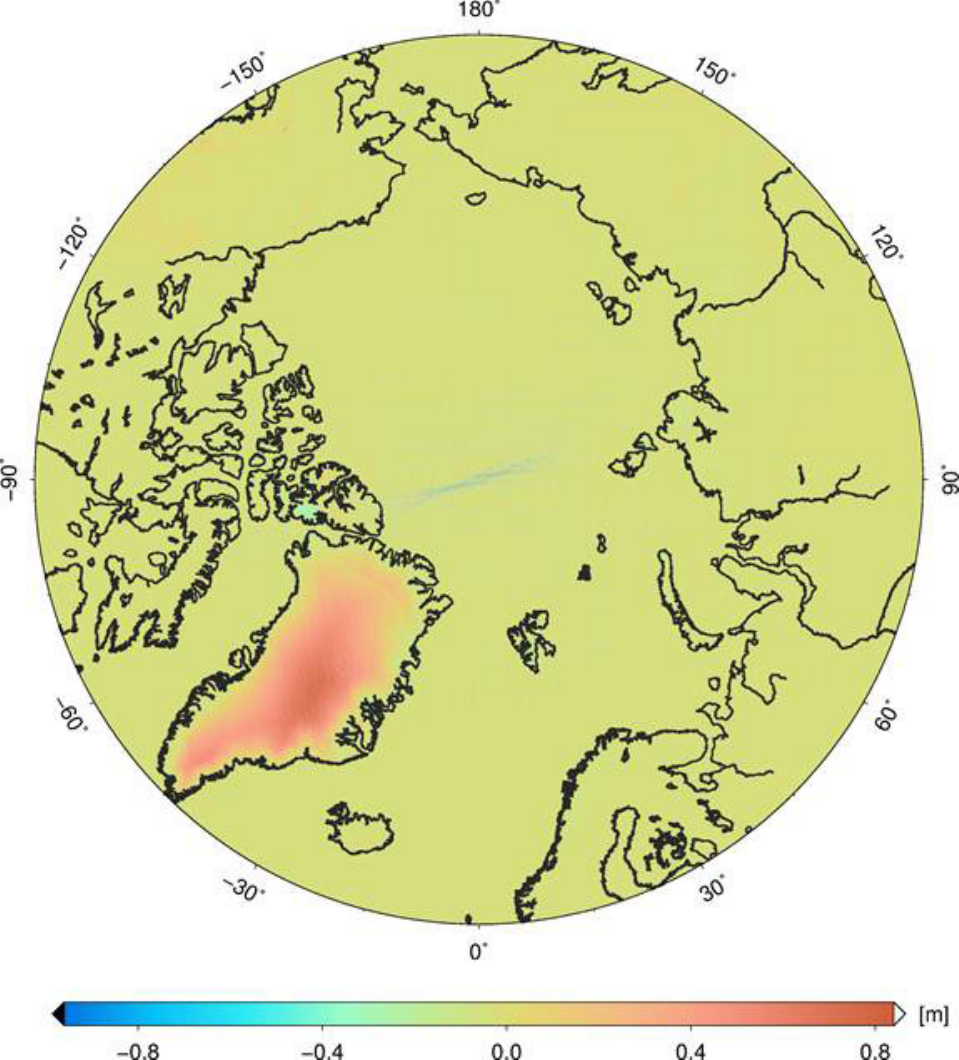}\\
  \caption{{\footnotesize
Differences between GGM2022 and E08GG in Arctic area
 }}\label{fig-Arctic-Diff-GGM2020-and-E08GG}
\end{figure}

\subsection{Comparison with GPS leveling data}\label{Comparison-with-GPS-leveling-data-subsection}

A reliable validation of GGM2022 relies on the GPS leveling benchmarks (GPSBMs). 

The globally available GPS/leveling benchmarks are shown by Figures\, \ref{fig-USA-GPS} to \ref{fig-westChina-GPS}, which respectively denote the USA, Australia, Europe, and western part of China. And comparative results between GGM2022 as well as E08GG and globally available GPS/leveling benchmarks are listed in tables \ref{tab-comparison-results}.

% \usepackage{multirow}
\begin{table*}
 \centering
\caption{Comparative results between GGM2022 as well as E08GG and globally available GPS/leveling benchmarks (Unit: m)}
\label{tab-comparison-results}
\begin{tabular}{ccccccc}
\hline
Region                                                                & Model  & Max   & Min    & Mean   & STD    & RMS   \\ \hline
\multirow{3}{*}{USA}                                                  & E08GG  & 0.184 & -1.321 & -0.514 & 0.3072 & 0.599 \\ \cline{2-7} 
                                                                     & GGM22$_-1.0$ $^*$ & 0.216 & -1.307 & -0.482 & 0.3071 & 0.571 \\ \cline{2-7} 
                                                                      & GGM22$_-2.0$ $^*$ & 0.216 & -1.311 & -0.485 & 0.3092 & 0.575 \\ \hline
\multirow{3}{*}{European}                                                   & E08GG  & 0.597 & -0.814 & -0.556 & 0.188 & 0.587 \\ \cline{2-7} 
                                                                      & GGM22$_-1.0$ $^*$ & 0.601`& -0.793& -0.554 & 0.189 & 0.585 \\ \cline{2-7} 
                                                                     & GGM22$_-2.0$ $^*$ & 0.600 & -0.890& -0.556 & 0.187 & 0.587 \\ \hline
\multirow{3}{*}{Australia}                                                  & E08GG  & 0.784  & -0.493& 0.141  & 0.197  & 0.243 \\ \cline{2-7} 
                                                                      & GGM22$_-1.0$ $^*$ & 0.809  & -0.498  & 0.144  & 0.198  & 0.245 \\ \cline{2-7} 
                                                                     & GGM22$_-2.0$ $^*$ & 0.806 & -0.498  & 0.143  & 0.198  & 0.244 \\ \hline
\multirow{3}{*}{\begin{tabular}[c]{@{}c@{}}Xinjiang\\ (China)\end{tabular}} & E08GG  & 0.319& -0.413 & -0.119 & 0.191  & 0.225 \\ \cline{2-7} 
                                                                      & GGM22$_-1.0$ $^*$ & 0.332 & -0.353& -0.115  & 0.189  & 0.221 \\ \cline{2-7} 
                                                                     & GGM22$_-2.0$ $^*$ & 0.128 & -0.543  & -0.118  & 0.178 & 0.214 \\ \hline
\end{tabular}

 {\tiny $^*$ GGM22$_-1.0$ $^*$ and GGM22$_-2.0$ $^*$ stand for GGM2020 based on CRUST1.0 and  CRUST2.0, respectively; E08GG$^*$ stands for EGM2008 Global Geoid}
\end{table*}

% \usepackage{multirow}
\begin{table*}
\centering
\caption{Comparative results between GGM2022 as well as E08GG and available GPS/leveling benchmarks at different elevations in the United States (Unit: m)}
\label{tab-comparison-of-diff-altitude}
\begin{tabular}{ccccccc}
\hline
Region                                                                                       & Model  & Max    & Min    & Mean   & STD   & RMS   \\ \hline
\multirow{3}{*}{\begin{tabular}[c]{@{}c@{}}USA\\ (low-lying, \textless{}300 m)\end{tabular}} & E08GG  & 0.183  & -1.299 & -0.359 & 0.272 & 0.451 \\ \cline{2-7} 
                                                                                             & GGM20$_-1.0$ $^*$ & 0.216  & -1.290 & -0.326 & 0.271 & 0.424 \\ \cline{2-7} 
                                                                                             & GGM20$_-2.0$ $^*$& 0.216  & -1.284 & -0.326 & 0.271 & 0.425 \\ \hline
\multirow{3}{*}{\begin{tabular}[c]{@{}c@{}}USA\\ (mountainous, 300-3700m)\end{tabular}}      & E08GG  & -0.144 & -1.299 & -0.709 & 0.239 & 0.748 \\ \cline{2-7} 
                                                                                             & GGM20$_-1.0$ $^*$ & -0.104 & -1.307 & -0.686 & 0.238 & 0.726 \\ \cline{2-7} 
                                                                                             & GGM20$_-2.0$ $^*$ & -0.116 & -1.312 & -0.706 & 0.235 & 0.744 \\ \hline
\end{tabular}

 {\tiny $^*$ GGM20$_-1.0$ $^*$ and GGM20$_-2.0$ $^*$ stand for GGM2020 based on CRUST1.0 and  CRUST2.0, respectively; E08GG$^*$ stands for EGM2008 Global Geoid}
\end{table*}

%In Australia we compare GGM2020 and E08GG with 2638 GPS points (Figure \ref{fig-AUS-GPS}). In Europe we compare GGM2020 and E08GG with 175 GPS points in north Europe (Figure \ref{fig-north-Europe}) . In China : We compare GGM2020 and E08GG with GPSBMs in all China (Figure \ref{fig-China-GPS}) and western part of China (Figure \ref{fig-westChina-GPS}). 

In USA, we compare GGM2022 and E08GG with 18972 GPS points (Figure \ref{fig-USA-GPS}). In the table \ref{tab-comparison-results}, we can see from the comparative results that the results based on Crust1.0 are better than E08GG in mean, standard deviation, and root-mean-square values.The results calculated by Crust2.0 are slightly worse than those calculated by Crust1.0, but the mean and root-mean-square values are still better than the E08GG results. In addition, we compared the calculations of GPS at different altitudes in the United States. In the table \ref{tab-comparison-of-diff-altitude} we can get the results. The improvement effect of GGM2022 is better in high-altitude areas than in low-altitude areas. This is because in the region with lower elevation, the terrain has less influence on the determination of geoid, so the difference between different density assumptions has less influence. However, in the higher altitude areas, the influence of topography cannot be ignored, and the difference brought by different density models will also become significant.
 
In Australia, we used 2638 GPS leveling points (Figure \ref{fig-AUS-GPS}) and in Europe we use146 GPS points in north Europe (Figure \ref{fig-north-Europe}).  As are shown in the table \ref{tab-comparison-results}, we can see that the results of GGM2022 is as well as E08GG in the two reguons.

In western part of China, we have 21 sparsely distributed GPSBMs in Xinjiang region (which are available at present) for the validation (see Figure \ref{fig-westChina-GPS}). Most of these points are located on rough mountains with maximum and mean elevations of 4936m and 1362m, respectively. The GGM2022 and E08GG are compared with the heights anomalies at the GPSBMs. Before the comparisons can be made, additional computations are performed to convert these geoid undulations into the corresponding height anomalies. Table \ref{tab-comparison-results} lists the statistical information for the comparisons. Based on Crust1.0, the differences between the height anomalies from the GGM2022 and the GPSBMs vary from -0.353 to 0.332 m, with a mean of -0.115m and based on Crust2.0, the differences between the height anomalies from the GGM2022 and the GPSBMs vary from -0.543 to 0.128 m, with a mean of -0.118 m. The standard deviations for the GGM2022(compared with 21 GPSBMs) are 18.9 cm and 17.8cm for Crust1.0 and Crust2.0 respectively while for the E08GG it is 19.1 cm, indicating that the GGM2022 is better than the E08GG in the Xinjiang region.  And for  root-mean-square values, GGM2022 is also better than E08GG.

\begin{figure}[!ht]
   \centering
  \includegraphics[scale=0.45]{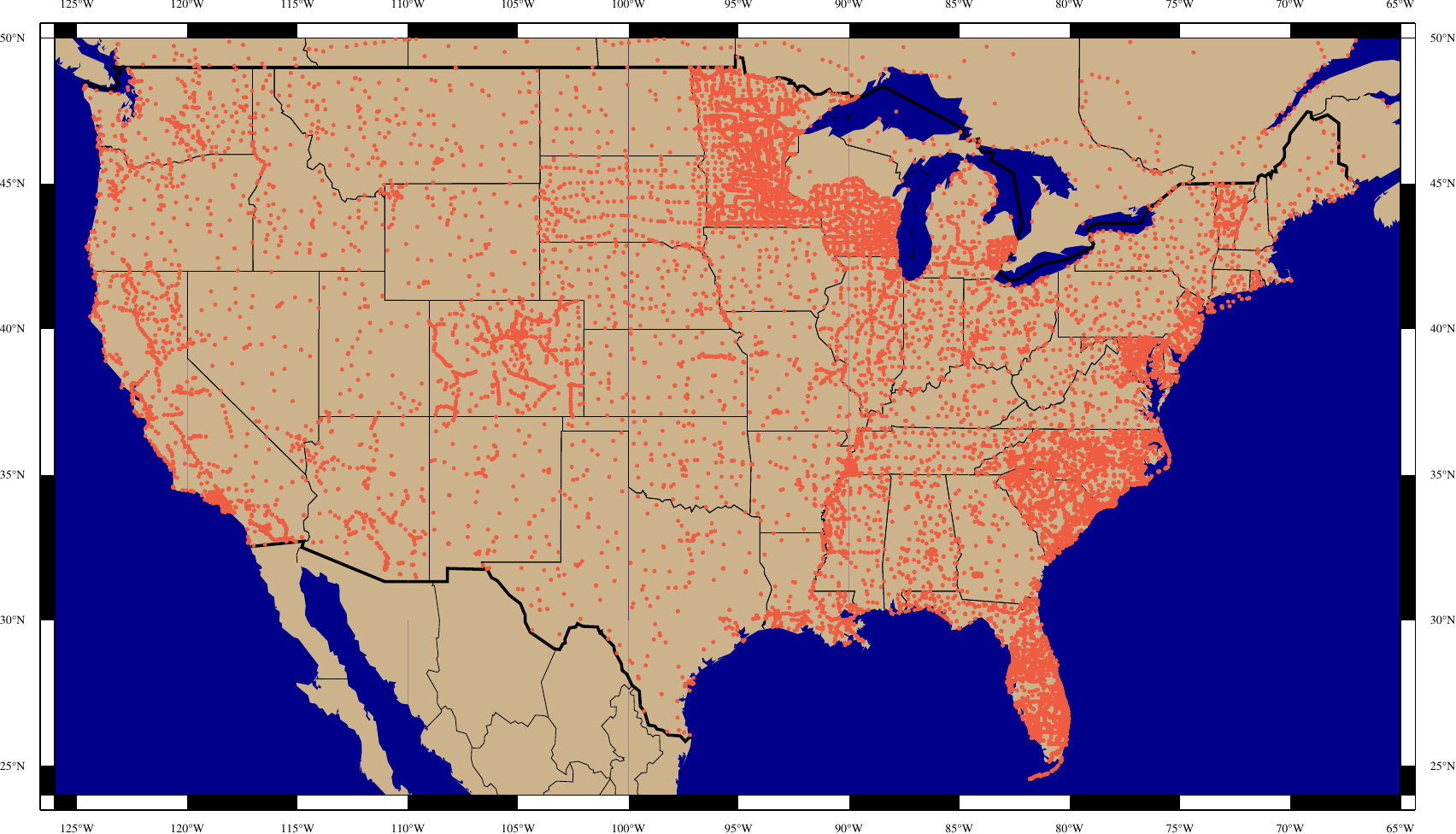}\\
  \caption{{\footnotesize
GPS leveling benchmarks in the USA (18972 points)
 }}\label{fig-USA-GPS}
\end{figure}

\vspace{3mm}

\begin{figure}[!ht]
   \centering
  \includegraphics[scale=0.75]{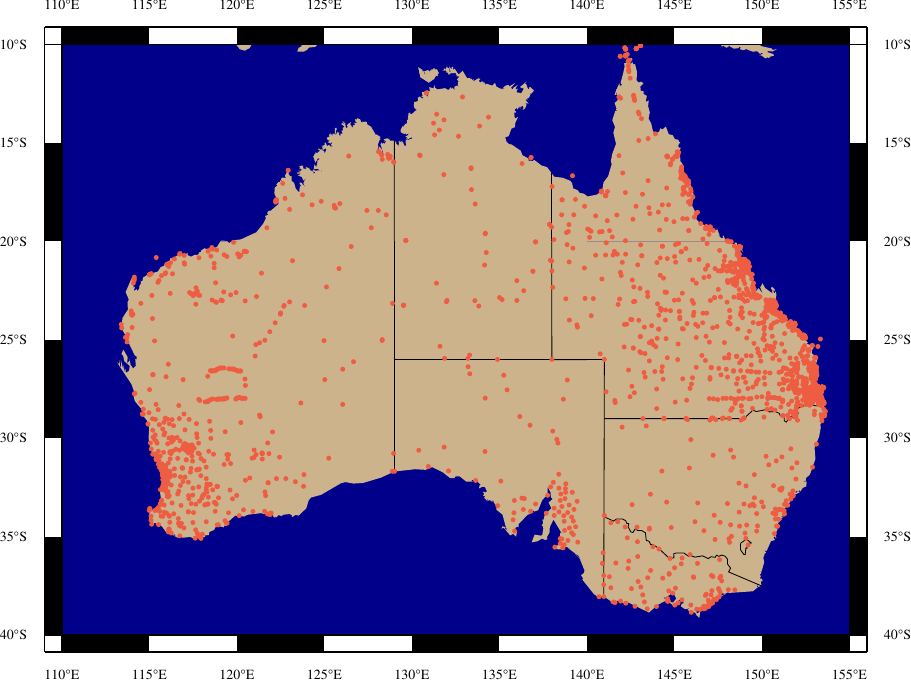}\\
  \caption{{\footnotesize
GPS leveling benchmarks in the Australia (2638 points)
 }}\label{fig-AUS-GPS}
\end{figure}

\vspace{3mm}

\begin{figure}[!ht]
   \centering
  \includegraphics[scale=0.75]{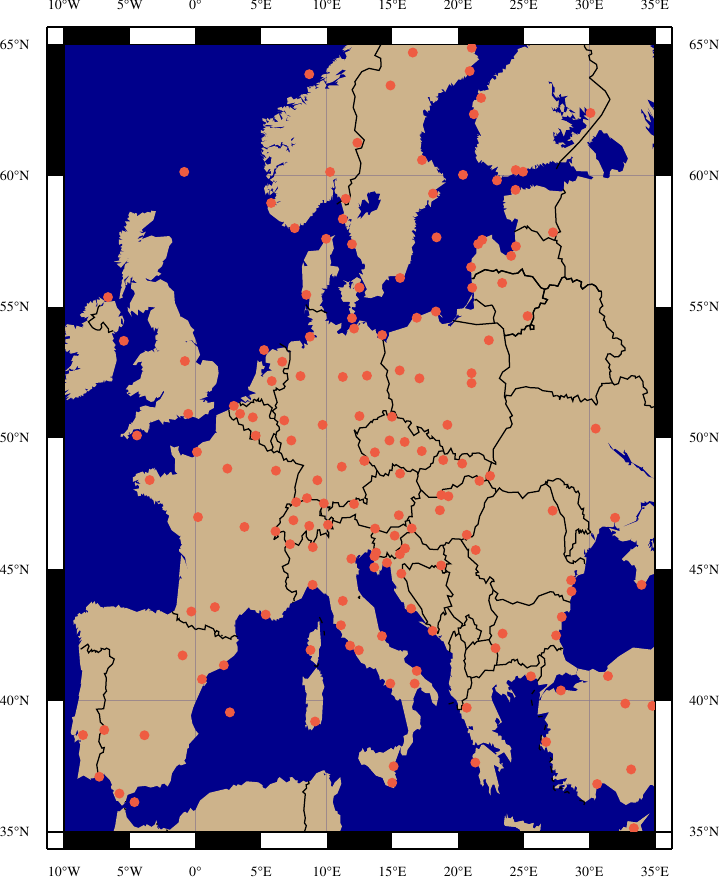}\\
  \caption{{\footnotesize
GPS leveling benchmarks in North Europe (146 points)
 }}\label{fig-north-Europe}
\end{figure}

\vspace{3mm}

%\begin{figure}[!h]
  % \centering
 % \includegraphics[scale=0.45]{fig-swi-Europe-eps-converted-to.pdf}\\
 % \caption{{\footnotesize
%GPS leveling benchmarks in Switzerland, Europe (?? points)
% }}\label{fig-swi-Europe}
%\end{figure}

%\vspace{3mm}

%\begin{figure}[!ht]
  
 % \centering
% \includegraphics[scale=0.45]{fig-China-GPS-eps-converted-to.pdf}\\
 % \caption{{\footnotesize
%GPS leveling benchmarks in China (649 points)
 %}}\label{fig-China-GPS}
%\end{figure}

%\vspace{3mm}%

\begin{figure}[!ht]
   \centering
  \includegraphics[scale=0.45]{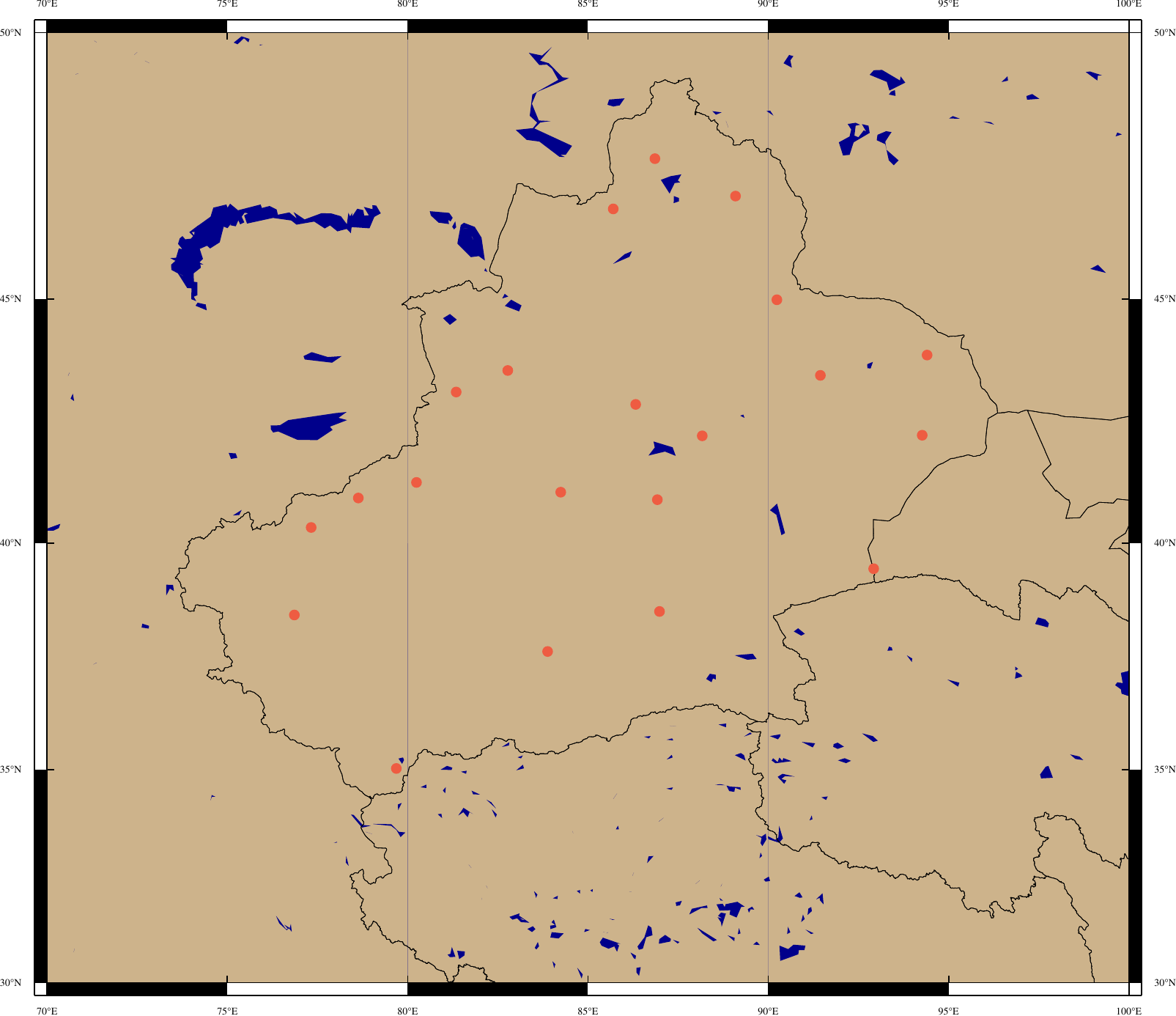}\\
  \caption{{\footnotesize
GPS leveling benchmarks in western part of China (21 points)
 }}\label{fig-westChina-GPS}
\end{figure}

\vspace{3mm}

\section{Conclusions and Discussions}\label{Conclusions-and-Discussions}

As an alternative approach for determining a global geoid, Shen method \citep{Shen2006} was proposed to overcome the drawbacks existing in Stokes method or Molodensky method. The main drawbacks of Stokes method lies in that it needs mass adjustment, removing the masses outside the geoid into the inside of the geoid, which influences the geoid determination. Though Molodensky method does not need mass adjustment, the determined geoid based on this method is a quasi-geoid, which is not an equi-geopotential surface. 

%Shen method relies on the DEM, external gravity field, and the density distribution of shallow layer. One may argue that if the density distribution outside the geoid is known very well, one can precisely determine the global geoid (say in the centimeter level) by using the conventional approaches (e.g.,Stokes approach or Molodensky approach). However, as pointed out in section 1, as well as pointed out by various authors, Stokes method may fail to complete this task because the masses outside the geoid should be removed inside the geoid, with the the pre-assumption that the geoid is known. However, if it is known, we needn't determine it at all. Of course iterative procedure could be applied, and assume that the results are closer to the real geoid with more and more iterations. 

Theoretically, using Shen method can determine a global geoid with certain resolution and accuracy that depend on the given datasets (models): (i) the Earth's surface with high resolution and accuracy (e.g., DEM); (ii) the density distribution of the shallow layer with good enough accuracy (e.g., refined crust density model CRUST$_-$re based on CRUST models and other additional information); (iii) the Earth's external gravitational potential field model with high resolution and accuracy (e.g., EGM2008 or EIGEN-6C4). This study and previous studies \citep{{Shen-and-Chao2008},{Shen-and-Han2012},{Shen-and-Han2013}}) suggest that Shen method is potential for determining a global/regional geoid. Using Shen method and based on EGM2008 model, DTM2006.0 model and DNSC08 MSS model, and CRUST$_-$re model, we provide here a $5 ^\prime \times 5 ^\prime $ global geoid model GGM2022. Evaluations show that GGM2022 fits the globally available GPS/leveling data better than EGM2008 global geoid in the USA, Europe and the western part of China . However, further improvements for global geoid determination can be executed concerning the following aspects. 

The EGM2020 may be released, and we can add it into our model GGM. Besides the external gravity field that we need, the accuracy of the global geoid depends mainly on the accuracy of the shallow mass layer model. To obtain more precise shallow mass layer model, we may replace ICE-5G \citep{Peltier2004} by an updated model ICE-6G$_-$C \citep{{Argus-et-al2014}, {Peltier-et-al2015}}, of which the latte is an improvement of the former. 

\noindent {\bf Acknowledgment:} This study is supported by  National Natural Science Foundation of China (NSFC) (grant Nos. 42030105, 42274011, 41721003, 41804012, 41874023) .

\vspace*{6mm}
\phantomsection \addcontentsline{toc}{section}{References}
\small
\bibliographystyle{agufull08}
\bibliography{shenwbEnglish2015Oct}

\end{document}